%% file: main_TMC.tex
\documentclass[10pt,journal,compsoc]{IEEEtran}

\input{Preamble}

\begin{document}
\title{Compressed Private
Aggregation for Scalable and Robust
Federated Learning \\ over
Massive Networks}

\author{Natalie~Lang,~
\IEEEmembership{Graduate~Student~Member,~IEEE,} Nir~Shlezinger,~\IEEEmembership{Senior~Member,~IEEE,}  Rafael~G.~L.~D'Oliveira,~\IEEEmembership{Member,~IEEE,} and~Salim~El~Rouayheb,~\IEEEmembership{Senior~Member,~IEEE}
\IEEEcompsocitemizethanks{
\IEEEcompsocthanksitem N. Lang and N. Shlezinger are with the School of  ECE, Ben-Gurion  University  of  the  Negev, Be’er-Sheva,  Israel (e-mails: langn@post.bgu.ac.il; nirshl@bgu.ac.il).
\IEEEcompsocthanksitem R. G. L. D'Oliveira is with the School of Mathematical and Statistical Sciences, Clemson University, SC (e-mail: rdolive@clemson.edu). 
\IEEEcompsocthanksitem S. El Rouayheb is with the Department of ECE, Rutgers University,	Piscataway, NJ (e-mail: salim.elrouayheb@rutgers.edu). 
}
\thanks{Parts of this work were presented in the 2023 IEEE International Conference on Acoustics Speech, and Signal Processing (ICASSP) as the paper \cite{lang2023cpa}.}}

\markboth{Journal of \LaTeX\ Class Files,~Vol.~14, No.~8, August~2015}%
{Shell \MakeLowercase{\textit{et al.}}: Bare Demo of IEEEtran.cls for Computer Society Journals}
\IEEEtitleabstractindextext{%
\begin{abstract}
\Ac{fl} is an emerging paradigm that allows a central server to train machine learning models using remote users' data.  Despite its growing popularity, \ac{fl} faces challenges in preserving the privacy of local datasets, its sensitivity to poisoning attacks by malicious users, and its communication overhead, especially in large-scale networks. These limitations are often individually mitigated by \ac{ldp} mechanisms, robust aggregation, compression, and user selection techniques, which typically come at the cost of accuracy. 
In this work, we present {\em \ac{cpa}}, allowing massive deployments to simultaneously communicate at extremely low bit rates while achieving privacy, anonymity, and resilience to malicious users. \ac{cpa} randomizes a codebook for compressing the data into a few bits using nested lattice quantizers, while ensuring anonymity and robustness, with a subsequent perturbation to hold \ac{ldp}. \ac{cpa}-aided \ac{fl} is proven to converge in the same asymptotic rate as \ac{fl} without privacy, compression, and robustness considerations, while satisfying both anonymity and \ac{ldp} requirements. These analytical properties are empirically confirmed in a numerical study, where we demonstrate the performance gains of \ac{cpa} compared with separate mechanisms for compression and privacy, as well as its robustness in mitigating the harmful effects of malicious users.
\end{abstract}
\acresetall 

}

\maketitle

\IEEEdisplaynontitleabstractindextext
\IEEEpeerreviewmaketitle

\section{Introduction}\label{sec:introduction}
\IEEEPARstart{T}{he} unprecedented success of deep learning highly relies on the availability of data, often gathered by edge devices such as mobile phones, sensors, and vehicles. As data may be private, there is a growing need to avoid leakage of private data while still being able to use it to train machine learning models. {\em\Ac{fl}}~\cite{mcmahan2017communication,kairouz2021advances, li2020federated, gafni2021federated} is an emerging paradigm for training a model on multiple edge devices, exploiting their computational capabilities \cite{chen2019deep}. 
\ac{fl} avoids directly sharing the users' data, as training is performed locally with periodic centralized aggregations of the models orchestrated by a server.

Learning in a federated manner is subject to several core challenges that are not encountered in traditional centralized \acl{ml} \cite{gafni2021federated,li2020federated}.
Despite the training being performed locally and not involving data sharing, it was recently shown that private information can be extracted and that the data can even be reconstructed from the exchanged models updates by model inversion attacks, if these are not properly protected \cite{zhu2020deep,zhao2020idlg,huang2021evaluating,yin2021see}. 
Furthermore, the fact that training is done on the users' side indicates that malicious users can affect the learned model by, for example, poisoning attacks~\cite{fang2020local, mothukuri2021survey}. 
Another prominent challenge stems from the repeated exchange of highly parameterized models between the server and the devices during the \ac{fl} procedure. As the communication links are possibly rate-limited channels, \ac{fl} can notably load the communication infrastructure, which, in turn, often results in considerable delays and degraded convergence \cite{karimireddy2020scaffold}, \cite{chen2021communication}.  
These usually become more dominant in large-scale \ac{fl} networks, causing significant overhead as well as yielding notable computational burden on the server side that recovers and aggregates the individual models, particularly when the number of users is huge, with possibly millions of participants. 

Various methods have been proposed to tackle the above challenges:
to preserve privacy, the \ac{ldp} framework is commonly adopted. \ac{ldp} quantifies privacy leakage of a single data sample when some function of the local datasets, e.g., a trained model, is publicly available \cite{kim2021federated}. 
\ac{ldp} can be boosted by corrupting the model updates with \ac{ppn} \cite{wei2020federated}, via splitting/shuffling \cite{sun2020ldp} or dimension selection \cite{liu2020fedsel}. 
An alternative privacy regime considered in \ac{fl} is \kano, which involves mechanisms that render certain features indistinguishable~\cite{yin2021comprehensive, kaissis2020secure}. 
Both \ac{ldp} and \kano mechanisms induce some level of perturbation that typically affects the learning procedure.
Considering the difficulty of dealing with unreliable and malicious users, this issue is typically addressed by Byzantine robust methods~\cite{vempaty2013distributed, yin2018byzantine, li2019rsa}. Such techniques have the servers use non-affine aggregation which reduces the sensitivity to outliers and thus limits the harmful effect of corrupted model updates; yet typically degrade performance in the absence of malicious users.  

The communication overhead of \ac{fl} is often relaxed by reducing the volume of model updates via lossy compression. 
This can be achieved by having each user transmit only part of its updates by sparsifying or sub-sampling \cite{han2020adaptive,konevcny2016federated, hardy2017distributed, aji2017sparse, alistarh2018convergence, lin2017deep}.  
An alternative approach discretizes the updates of the model through quantization, so that it is conveyed using a small number of bits \cite{alistarh2017qsgd,reisizadeh2020fedpaq,bernstein2018signsgd,shlezinger2020uveqfed, horvath2019natural}. 
Scalability is typically enabled by limiting the number of participating devices through user selection~\cite{chen2021communication}. These methods determine which of the users participate in each round of training, taking into account the individual constraints of computation and communication resources \cite{nishio2019client}, as well as the magnitude of local updates~\cite{chen2021communication}. 

Recent studies consider both the challenges of compression and privacy in \ac{fl}. 
The works \cite{lyu2021dp, zhang2022leveraging} propose to quantize the local gradient with a differentially private 1-bit compressor; \cite{amiri2021compressive,lang2022joint},\cite{chaudhuri2022privacy} employ probabilistic quantizers to achieve compression in a manner that also enhances privacy, so that the incorporation of a dedicated \ac{ppn} can result in the compressed representation obeying established multivariate \ac{ldp} mechanisms~\cite{lang2022joint}. 
All these schemes have the server separately recover the model updates for each user and then aggregate via conventional averaging. Thus, they are neither inherently scalable to suit massive systems and tolerate large groups of colluding users, nor account for robustness considerations. 

\begin{table*}
\caption{Comparison between \ac{cpa} and existing \ac{fl} studies}
\label{tbl:existing_FL_literature}
\centering
\begin{adjustbox}{width=\textwidth} 
\begin{tabular}{|c||c|c|c|c|c|}
\hline
 & \bfseries Compression & \multicolumn{2}{|c|}{ \bfseries Proven Privacy} & \bfseries Security & \bfseries Scalability \\
\cline{2-6}
&  & \bfseries \ac{ldp} & \bfseries Anonymity &  & \\
\hline\hline
Byzantine robust aggregation, e.g., \cite{vempaty2013distributed, yin2018byzantine, li2019rsa} & \xmark & \multicolumn{2}{c|}{\xmark} & \cmark & \xmark\\
User selection, e.g., \cite{chen2021communication, nishio2019client} & \xmark & \multicolumn{2}{c|}{\xmark} & \xmark & \cmark\\
Model updates compression, e.g., \cite{han2020adaptive,konevcny2016federated, hardy2017distributed, aji2017sparse, alistarh2018convergence, lin2017deep,alistarh2017qsgd,reisizadeh2020fedpaq,bernstein2018signsgd,shlezinger2020uveqfed, horvath2019natural} & \cmark &  \multicolumn{2}{c|}{\xmark} & \xmark & \xmark  \\
Privacy enhancement, e.g., \cite{kim2021federated,wei2020federated,sun2020ldp,liu2020fedsel} & \xmark & \cmark & \xmark & \xmark & \xmark\\
Joint compression \& privacy, e.g., \cite{lyu2021dp, zhang2022leveraging, amiri2021compressive,lang2022joint, chaudhuri2022privacy} & \cmark &  \cmark & \xmark & \xmark & \xmark  \\
\ac{cpa}  & \cmark & \cmark & \cmark & \cmark & \cmark\\
\hline
\end{tabular}
\end{adjustbox}
\end{table*}

In this work, we present a novel privacy preserving scheme designed for robust large-scale \ac{fl}. The method, coined {\em \ac{cpa}}, dramatically reduces communications by conveying model updates via messages of only a few bits, while providing $k$-anonymity and \ac{ldp}, as well as limiting the individual contribution of each user to increase robustness to malicious users.
Unlike existing \ac{fl} techniques, \ac{cpa} jointly provides compression, proven privacy, inherent scalability, and empirically observed Byzantine robustness, without limiting learning capabilities, as summarized in Table~\ref{tbl:existing_FL_literature}.
It is inspired by \acl{pmga} \cite{naim2022private} and geo-indistinguishability \cite{thesmar2021cabdriver}, which involve settings that fundamentally differ from \ac{fl} in their task, yet inherently employ massive systems where scalability and robustness are key factors. 

We design \ac{cpa} by leveraging nested lattice quantizers~\cite{abdi2019nested} combined with random codebooks to encode the set of model updates into few bits at each user. The discretizing operation of the quantizers is exploited to provide anonymity, and is then further perturbed by incorporating an established \ac{rr} mechanism. We analytically show that the resulting representation conveyed by each user rigorously holds both \kano and \ac{ldp} guarantees, and empirically demonstrate that it utterly limits each user's influence and leads to robustness from different forms of corrupted models.The conveyed few-bit representations are aggregated by the server via a decoding procedure, translating the received bits from all different users into an empirical discrete histogram over the model update values. 

The aggregated mean of this histogram is shown to converge into the averaged global trained model, yielding the desired updated global model in each \ac{fl} iteration. By doing so, the server does not reconstruct the individual model updates, which, when combined with the few-bit communication involved in \ac{cpa}, notably facilitates the participation of numerous users and supports scalability. Furthermore, we systematically show that the distortion in the resulting aggregated model compared to vanilla \ac{fl} (without communication, privacy, or security considerations) decreases as the number of users increases, and that the resulting model converges in the same asymptotic order as vanilla \ac{fl}. These theoretical findings are numerically demonstrated in our experimental study. There, we evaluate \ac{cpa} for learning several different image classification models, showing that its overall distortion is reduced compared to conventional methodologies for private compressed \ac{fl}, and that this reduced distortion is translated into an improved performance of the learned model.

Our main contributions are summarized as follows: 
\begin{itemize}
    \item {\bf Novel scalable aggregation technique:} \ac{cpa} presents a joint design of probabilistic model quantization and users `voting' for private aggregation. The perturbation introduced therein to meet \ac{ldp}, is mitigated not by the conventional \ac{fa} but rather by a unique reconstruction of discrete histograms, having the server avoids recovering the individual updates for each user. While this allows applicability over large-scale \ac{fl} systems, it also guarantees \kano by design. 
    \item {\bf Byzantine robustness following user's low-influence:} \ac{cpa} exploits the high-dimensional structure of the model updates through (possibly high-rate) lattice quantization but still dramatically reduces the conventional \ac{fl} communication overhead. This follows since the users transmit at most $B$ bits per sample, which inherently limits their influence on the final model and allows the training to be Byzantine robust against erroneous adversarial users and poisoning attacks.
    \item {\bf Theoretical and experimental evaluation:}
    The ideas introduced in \ac{cpa} draw inspiration from previous studies in different domains on model compression and private aggregation. The novelty and contribution of \ac{cpa} relies on coupling and fusing these parallel domains in a noise-controllable manner. The ability to learn reliably in large scale networks systematically exemplified for \ac{cpa} in both analytical and extensive numerical analysis.
\end{itemize}

The remainder of this paper is organized as follows: Section~\ref{sec:prelim} briefly reviews the \ac{fl} system model and the relevant preliminaries. \ac{cpa} is presented in Section~\ref{sec:method}, while Section~\ref{sec:analysis} theoretically analyzes its privacy guarantees and convergence profile. In Section~\ref{sec:experiments} we numerically evaluate \ac{cpa}. Finally, Section~\ref{sec:conclusions} provides concluding remarks.

Notations: throughout this paper, we use boldface lowercase letters for vectors, e.g., $\vx$, boldface uppercase letters for matrices, e.g., $\vX$, and calligraphic letters for sets, e.g., $\cX$. The stochastic expectation, variance, and $\ell_2$ norm are denoted by $\E[\cdot]$,  $\var(\cdot)$, and $\|\cdot\|$, respectively, while $\Z$ and $\R$ are the sets of integer and real numbers, respectively. 

\section{System Model and Preliminaries}\label{sec:prelim}
In this section we present the system model of bit-constrained and private \ac{fl}. We begin by recalling some relevant basics in \ac{fl} and quantization in Subsections~\ref{subsec:FL}-\ref{subsec:quantization} respectively. We then review the privacy preliminaries in Subsection~\ref{subsec:privacy}, and formulate our problem in Subsection~\ref{subsec:prebelm_description}.

\subsection{Federated Learning}\label{subsec:FL}
In \ac{fl}, a server trains a machine learning model parameterized by  $\vw\in \R^\dimension$ using data available at a group of $K$ users indexed by $1,\ldots, K$. These datasets, denoted $\cD_1,\dots, \cD_K$, are assumed to be private. Thus, as opposed to conventional centralized learning where the server can use $\cD=\bigcup_{\user=1}^K \cD_\user$ to train $\vw$, in \ac{fl} the users cannot share their data with the server.  
Let $F_\user(\vw)$ be the empirical risk of a model $\vw$ evaluated over the dataset $\cD_\user$. The training goal is  to recover the $\dimension\times 1$ optimal weights vector $\vw^{\rm opt}$ satisfying
\begin{equation}\label{eq:w_opt}
    \vw^{\rm opt} = \argmin_{\vw} \left\{F(\vw)\triangleq \frac{1}{K}\sum_{\user=1}^K  F_\user\left(\vw\right)\right\}.
\end{equation}
Generally speaking, \ac{fl} involves the distribution of a global model to the users. Each user locally trains this  model using its own data and sends back the model update~\cite{gafni2021federated}. Therefore, users do not directly expose their private data, as training is performed locally. The server then aggregates the models into an updated global model and the procedure repeats iteratively. 

Arguably, the most common \ac{fl} scheme is {\em \ac{fa}}~\cite{mcmahan2017communication}, where the global model is updated by averaging the local models. Letting $\vw_t$ denote the global parameters vector available at the server at time step $t$, the server shares $\vw_t$ with the users, who each performs $\tau$ training iterations using its local $\cD_\user$ to update $\vw_t$ into $\vw^\user_{t+\tau}$. 
Typically, the information conveyed from the users to the server is not the model weights, i.e., $\vw^\user_{t+\tau}$, but the updates to the model generated in the current round, i.e., $\vh^\user_{t+\tau} \triangleq \vw^\user_{t+\tau} - \vw_t$. As the server knows $\vw_t$, it recovers $\vw_{t+\tau}$ from the difference $\vw^\user_{t+\tau} - \vw_t$. 
The server in turn sets the global model to be 
\begin{align}\label{eq:fed_avg}
    \vw_{t+\tau} \triangleq \vw_t + \frac{1}{K}\sum_{\user=1}^K  \vh^\user_{t+\tau}=\frac{1}{K}\sum_{\user=1}^K  \vw^\user_{t+\tau},
\end{align}
where it is assumed for simplicity that all $K$ users participate in each \ac{fl} round. The updated global model is again distributed to the users and the learning procedure continues. 

When the local optimization at the users' side is carried out using \ac{sgd}, then \ac{fa} applies the {\em \lsgd} method~\cite{stich2018local}. In this case, each user of index $\user$ sets $\vw_t^\user = \vw_t$, and updates its local model via 
\begin{align}\label{eq:local_sgd}
    \vw^\user_{t+1} \xleftarrow{}  \vw^\user_t -\eta_t \nabla F^{\sample}_\user\left(\vw^\user_t\right),
\end{align}
where $\sample$ is the sample index chosen uniformly from $\cD_\user$, $\eta_t$ is the learning rate, and $F^{\sample}_\user(\cdot)$ is the empirical risk computed using the $\sample$-th sample in $\cD_\user$. 
As sharing $\vw^\user_{t+\tau}$ can possibly load the communication network and leak private information, it motivates the integration of quantization and privacy enhancement techniques, discussed below.

\subsection{Quantization Preliminaries}\label{subsec:quantization}
Vector quantization is the encoding of a set of continuous-amplitude quantities into a finite-bit representation~\cite{gray1998quantization}. 
Vector quantizers which are invariant of the underlying distribution  of the vector to be quantized are referred to as {\em universal vector quantizers}; a leading approach to implement such quantizers is based on lattice quantization~\cite{zamir1992universal}:
\begin{definition}[Lattice Quantizer]\label{def:lattice_quantizer}
A lattice quantizer of dimension $L \in \Z^+$ and  generator matrix $\vG\ \in \R^{L\times L}$ maps  $\vx\in \R^L$ into a discrete representation $Q_\cL(\vx)$ by selecting the nearest point in the lattice  $\cL \triangleq \{ \vG\vl: \vl\in\Z^L\}$, i.e., 
\begin{equation}
    Q_\cL(\vx) = \argmin_{\vz \in \cL}\|\vx-\vz\|. 
\end{equation}
\end{definition}
To apply $Q_\cL$ to a vector $\vx \in \R^{ML}$, it is divided into ${[\vx_1,\dots,\vx_M]}^T$, and each sub-vector is quantized separately. 

A lattice $\cL$ partitions $\R^L$ into cells centered around the lattice points, where the basic cell is $\cP_0=\{\vx:Q_\cL(\vx)=\boldsymbol 0\}$. The number of lattice points in $\cL$ is countable but infinite. Thus, to obtain a finite-bit representation, it is common to restrict $\cL$ to include only points in a given sphere of radius $\gamma$, $\cL_\gamma$,
and the number of lattice points, $|\cL_\gamma|$, dictates the number of bits per sample -- 
$R\triangleq\frac{1}{L}\log_2(|\cL_\gamma|)$. 
An event in which the input to the lattice quantizer does not reside in this sphere is referred to as {\em overloading}, from which quantizers are typically designed to avoid \cite{gray1998quantization}. 
In the special case of $L=1$ with $\vG=\Delta_{\rm Q} >0$, $Q_\cL(\cdot)$ specializes conventional scalar uniform quantization $Q(\cdot)$.
\begin{definition}[Uniform Quantizer]\label{def:scalar_uniform_Q}
A mid-tread scalar uniform quantizer with support $\gamma$ and spacing $\Delta_{\rm Q}$ is defined as
\begin{align}\label{eq:scalar_uniform_Q}
    Q(x) = 
    \begin{cases}			
    \Delta_{\rm Q}\left\lfloor\frac{x}{\Delta_{\rm Q}} + \frac{1}{2}\right\rfloor & \text{if } x<\abs{\gamma},\\
    \sign(x)\cdot \left(\gamma -\frac{1}{2}\Delta_{\rm Q}\right)  & \text{otherwise}
    \end{cases}
\end{align}
where $R=\log_2 \left(2\gamma / \Delta_{\rm Q}\right)$ bits are used to represent $x$.
\end{definition}

The formulation of lattice and uniform quantizers in Defs.~\ref{def:lattice_quantizer}-\ref{def:scalar_uniform_Q} gives rise to two extensions, which are adopted in the sequel. The first is {\em Probabilistic quantization}, which converts the quantizers to implement stochastic mapping. A conventional probabilistic quantization technique uses \ac{dq}, which applies $ Q_\cL$ to a noisy version of the input~\cite{lipshitz1992quantization, gray1993dithered}. When the added noise is uniformly distributed over $\cP_0$ and the quantizer is not overloaded, the resulting distortion becomes an i.i.d. stochastic process~\cite{gray1993dithered,zamir1996lattice}.

\begin{figure}
\centering    
\includegraphics[width=0.65\columnwidth]{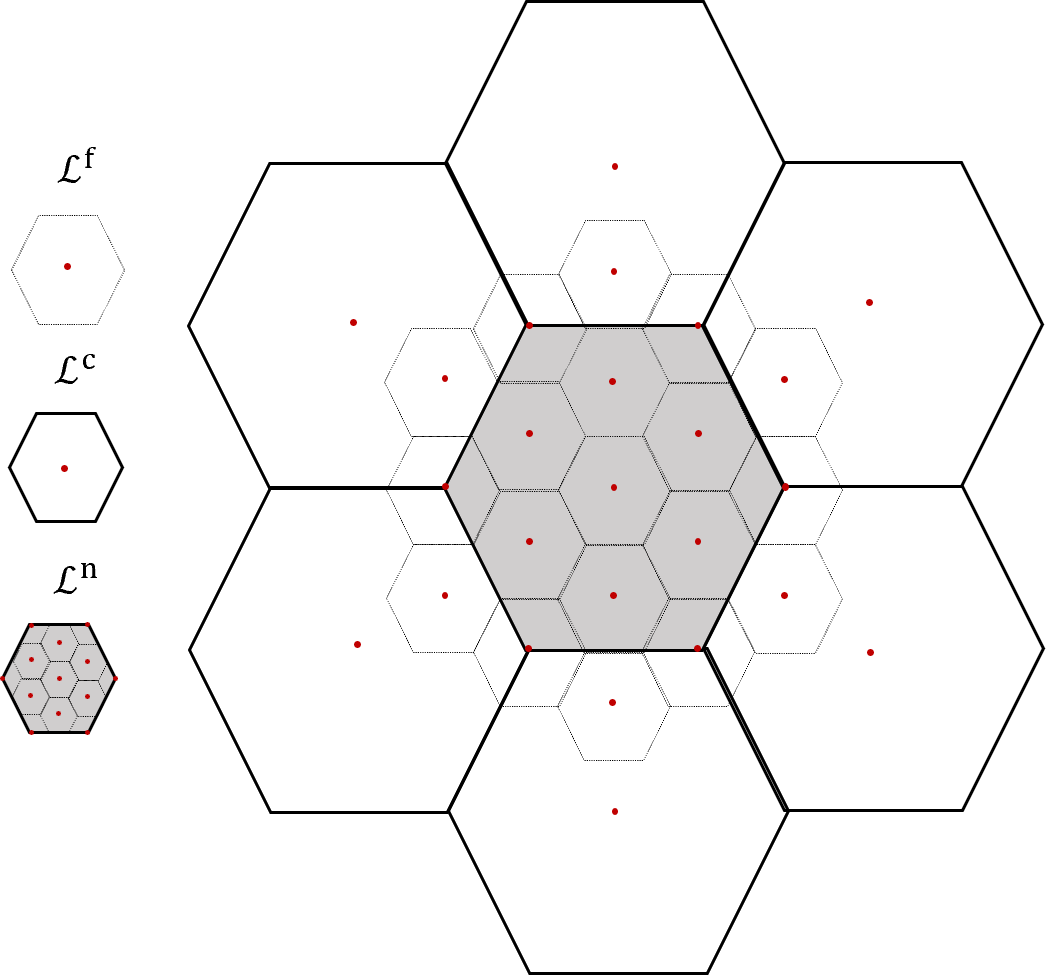}

\vspace{3mm}

\includegraphics[width=\columnwidth]
{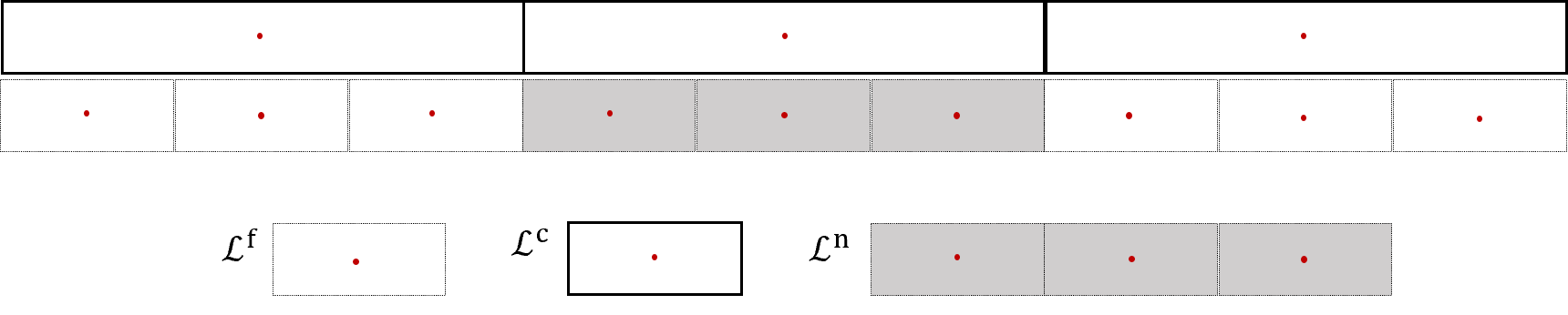}
    \caption{Nested self-similar lattice quantizers for $L=2$ (top) and $L=1$ (bottom).}
    \label{fig:nestedQ}
\end{figure}

The second extension of lattice quantizers is \nq, which implements increased resolution quantization using multiple low-resolution quantizers. For simplicity, we now focus on \nq with two quantizers, as visualized in Fig.~\ref{fig:nestedQ}. The formal definition of a two-stage nested lattice quantizer is as follows~\cite{zamir2002nested}.

\begin{definition}[Nested Lattice Codebook~\cite{polyanskiy2014lecture}]\label{def:nestedLQ}
A lattice $\coarse$ is said to be nested in $\fine$ if $\coarse \subset \fine$.  
Let $\coarseP$ denote the basic lattice cell of $\coarse$, 
then a nested lattice codebook $\cL^{\rm n}$ based on the nested lattice pair $\coarse \subset \fine$ is defined as
\begin{equation}
    \nested\triangleq \fine \cap \coarseP.
\end{equation} 
\end{definition}

A nested formulation allows quantizing with the fine lattice quantizer ($\fine$) using the nested ($\nested$) and the coarse ($\coarse$) ones. In particular, one can quantize $\vx \in \R^L$ by computing $Q_{\fine}(\vx) = Q_{\coarse}(\vx)+Q_{\cL^{\rm n}}\big(\vx - Q_{\coarse}(\vx)\big)$. 
Nested lattice quantizers naturally specialize multi-bit scalar uniform quantization~\cite{abdi2019nested}, where the nesting condition implies that the quantization spacing of the coarse quantizer must be an integer multiple of the corresponding spacing of the fine quantizer.
While Def.~\ref{def:nestedLQ} is given for a two-stage quantizer, i.e., $\fine$ is implemented using two quantizers, it can be recursively extended into multiple stages by quantizing over $\nested$ in a nested fashion. 

\subsection{Privacy Preliminaries}\label{subsec:privacy}
Privacy in settings involving queries between users and a server is commonly quantified in terms of \ac{dp}~\cite{yang2017survey, abowd2018us} and \ac{ldp}~\cite{kasiviswanathan2011can, wang2020federated}. 
While both provide users with privacy guarantees from untruthful adversaries, the latter does not assume a trusted third-party server, and is thus commonly adopted in \ac{fl}~\cite{wang2020federated, zhao2020local, xiong2020comprehensive, kim2021federated, sun2020ldp, liu2020fedsel, lang2022joint}. Therefore, we consider \ac{ldp} in this work, defined below.
\begin{definition}[$\varepsilon$-\ac{ldp}~\cite{wang2020comprehensive}]\label{def:LDP}
A randomized  mechanism $\cM$ satisfies $\varepsilon$-\ac{ldp} if for any pairs of input values $v,v'$ in the domain of $\cM$ and for any possible output $y$, it holds that
\begin{align}\label{eq:LDP}
    \Pr [\cM(v)=y] \leq e^\varepsilon \Pr [\cM(v')=y].
\end{align}
\end{definition}
We note that a smaller $\varepsilon$ means greater protection of privacy. Def.~\ref{def:LDP} implies that privacy can be achieved by stochasticity: if two different inputs are probable (up to some privacy budget) to be associated with the same algorithm output, then privacy is preserved, as each sample is not uniquely distinguishable. 

For continuous quantities, common mechanisms that achieve $\varepsilon$-\ac{ldp} are widely based on perturbation with \acf{ppn}, e.g., Laplacian or multivariate $t$~\cite{reimherr2019elliptical}. The \ac{ppn} distribution parameters set to  meet the \ac{ldp} privacy level $\varepsilon$.  
For private binary queries, a principle method for achieving $\varepsilon$-\ac{ldp} is the {\em \ac{rr} mechanism} \cite{warner1965randomized}. 
In \ac{rr}, a user who possesses a private bit transmits it correctly with probability $p>1/2$.
By \eqref{eq:LDP}, it can be shown that \ac{rr} satisfies $\log \left( \frac{p}{1-p} \right)$-\ac{ldp}~\cite{wang2020comprehensive} 
and can be viewed as a \ac{ppn} mechanism. 

Although \ac{ldp} is a preferable privacy measure, it is often guaranteed by the introduction of a dominant \ac{ppn} perturbations. Alternative privacy measures, which are not inherently bundled with stochasticity, are based on anonymization \cite{wang2020comprehensive} such as \kano~\cite{sweeney2002k}:
\begin{definition}[\kano~\cite{sweeney2002k}]\label{def:kAno}
A deterministic mechanism $\cM$ holds \kano if for every input  $v$ in the domain of $\cM$ there are at least $k-1$ different inputs  $\{v'_i\}_{i=1}^{k-1}$ satisfying 
\begin{align}\label{eq:kAno_def}
    \cM(v) = \cM(v'_i), \qquad \forall i\in\{1,\ldots,k-1\}.
\end{align}
\end{definition}
\noindent
If $\cM$ satisfies \kano, any observer of $\cM$'s output is unable to discriminate between at least $k$ possible inputs.

\subsection{Problem Description}\label{subsec:prebelm_description}
\subsubsection{Threat Model}
\Ac{fl} was shown to be exploitable by adversaries, with various possible attacks and threat models~\cite{usynin2021adversarial}. 
Here, we focus on two types of threats. The first is privacy attacks, i.e., algorithms that reconstruct the raw original private data, based on unintentional information leakage regarding the data or the machine learning model, being a unique characteristic of \ac{fl}. Inspired by \cite{geiping2020inverting}, we investigate an {\em honest-but-curious server} with the goal of uncovering the users' data. The attacker is allowed to separately store and process updates transmitted by individual users, but may not interfere with the learning algorithm. The attacker may not modify the model architecture nor may it send malicious global parameters that do not represent the actual global learned model.  

An additional threat considered is that of {\em adversarial participants}, often assumed in Byzantine robust \ac{fl}~\cite{yin2018byzantine}. Under this model, an unknown subset of the participating users may convey corrupted model updates,  
via poisoning attacks~\cite{fang2020local}. The identity of the unreliable users is not known to the server nor to the remaining reliable participants. 

\subsubsection{Problem Formulation}
Our goal is to design a global aggregation mechanism \cite{gafni2021federated} for \ac{fl} that provides privacy guarantees, compression, robustness, and is scalable. In particular, we are interested in obtaining a mapping $\vh_t^\user \mapsto \vw_t$ of the local updates at the $\user$-th user into the global model available at the server. The scheme must be:
\begin{enumerate}[label={\em R\arabic*},leftmargin=*]
\itemsep0em 
    \item \label{itm:ldp} {\em Private}: holding \kano and $\varepsilon$-\ac{ldp} with respect to the private datasets $\{\cD_\user\}$, for a given anonymity degree $k$ and privacy budget $\varepsilon$, respectively.
    \item \label{itm:rate} {\em Compressed}: communications to the server should involve at most $B$ bits per sample.
    \item \label{itm:universal} {\em Universal}: invariant to the distribution of $\vh_t^\user$.     
    \item \label{itm:robustness} {\em Robust}: resilient to adversarial participants and tolerate a large group of colluding users.
    \item \label{itm:scalability} {\em Scalable}: operable with possibly millions of participants.
\end{enumerate}

In \ref{itm:ldp} we focus on achieving \ac{ldp} in each round of communication. One can use a per-round privacy level to formulate an overall privacy guarantee after a given amount of rounds via the composition theorem~\cite[Thm. III.1.]{dwork2010boosting}, as the overall privacy level after $T$ rounds is at most $T \cdot \varepsilon$. Nevertheless, recent work has shown that by additional processing, a per-round privacy level can be translated into a multi-round one, obtaining an overall privacy budget depending on $\varepsilon$ that does not linearly grow with the number of rounds~\cite{sun2020ldp}. For these reasons, we formulate our privacy budget as in~\ref{itm:ldp}.

Evidently, requirements \ref{itm:ldp}-\ref{itm:universal} can be satisfied by first perturbing the data to meet \ref{itm:ldp}, followed by universal quantization to satisfy \ref{itm:rate}-\ref{itm:universal}, as both techniques are invariant to the distribution of $\vh_t^\user$. 
However, the server decoding in these separate schemes requires individual reconstruction, which may result in violating \ref{itm:scalability} while not accounting for \ref{itm:robustness}. Furthermore, both privacy and quantization can be modeled as corrupting the model updates, 
and thus using separate mechanisms may result in an overall noise which degrades the accuracy of the trained model beyond that needed to meet \ref{itm:ldp}-\ref{itm:universal}. These observations motivate a joint design tailored for \ac{fl}, studied next.

\begin{figure*}
     \centering              
     \includegraphics[width=\textwidth]{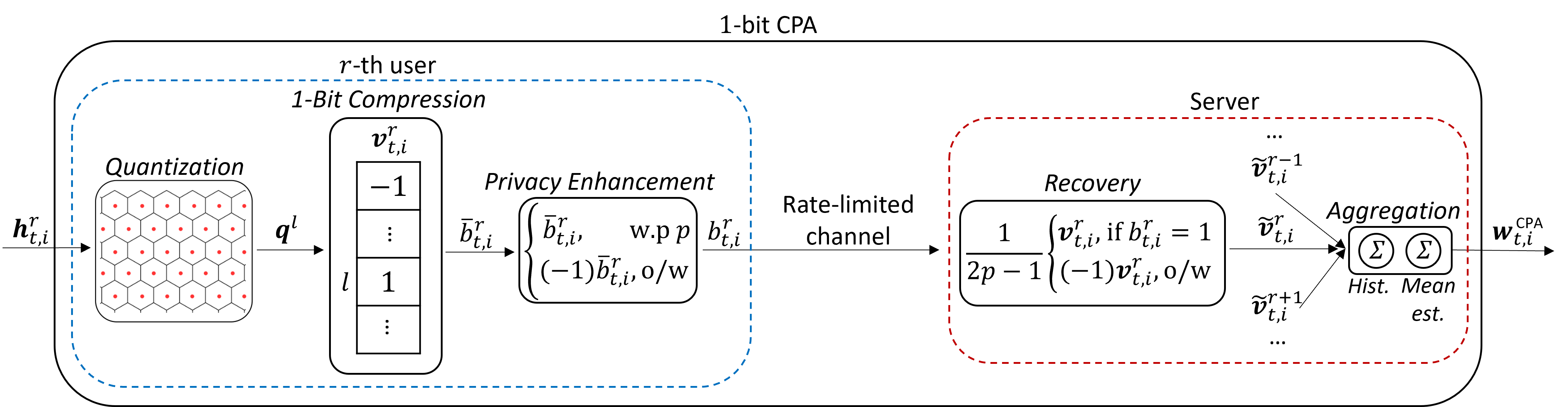}
    \caption{Overview of \ac{cpa}. The left dashed box represents the $\user$-th user encoding while the right describes the server decoding.}
    \label{fig:cpa}
\end{figure*}

\section{Compressed Private Aggregation}\label{sec:method}
In this section we introduce \ac{cpa}, deriving its basic steps for 1-bit messages in Subsection~\ref{subsec:cpa}, and its extension to multi-bit messages via nested quantization in Subsection~\ref{subsec:nested_cpa}. Then,  we 
provide a discussion in Subsection~\ref{subsec:discussion}. 

\subsection{1-Bit CPA}\label{subsec:cpa}
We design \ac{cpa} based on \ref{itm:ldp}-\ref{itm:scalability} by extending the recent schemes of \cite{naim2022private} and \cite{thesmar2021cabdriver} to \ac{fl} settings. Broadly speaking, \ac{cpa} leverages the repeated communications of \ac{fl} to generate a random codebook and encode the data with the help of an $L$ dimension lattice quantizer (holding \ref{itm:universal}). The generated code enables the transmission of a set of $L$ model update entries with a single bit, i.e., $B=\frac{1}{L}$ (satisfying \ref{itm:rate}), which guarantees \kano of the data. We then support \ac{ldp} by applying \ac{rr} to the transferred bits (satisfying \ref{itm:ldp}). In the decoding procedure, the received bits are translated into an empirical histogram over the model update values, rather than recovering each model update separately (holding \ref{itm:scalability}). The aggregated mean over this histogram converges into the \ac{fa} trained model, inherently limiting the influence of potential malicious participating users, as they can, at most, flip one bit (assuring \ref{itm:robustness}). These steps, illustrated in Fig.~\ref{fig:cpa} and summarized as Algorithm~\ref{alg:cpa}, are described below in detail.

\subsubsection{Initialization}
To initialize \ac{cpa}, the privacy parameters $k$ and $\varepsilon$ are set, and the compression lattice $\cL$ is determined, i.e., fixing the dimension of the lattice $L$, its generator matrix $\vG$, radius $\gamma$, and rate $R$ \cite[Ch. 2]{conway2013sphere}. We allow the lattice to change over the \ac{fl} rounds, and thus denote it by $\cL_t$. The motivation to do so is to allow the quantizer to adapt its mapping along the \ac{fl} procedure, and particularly by gradually decreasing the dynamic range over time to better represent the model updates whose magnitude typically decreases as \ac{fl} approaches converges. Additionally, a common seed $s_\user$ is shared between each user and the server. This can be provided by the user along with the initial sharing of updates in the \ac{fl} procedure, as done in, e.g.,~\cite{shlezinger2020uveqfed}. 

\subsubsection{Encoding}\label{subsubsec:encoding} 
The \ac{cpa} procedure is carried out on each \ac{fl} global aggregation round. Therefore, we describe it for a given time step $t$, in which the users have updated their local models. Since the encoding is identical for all users, we focus on the $\user$-th user, who is ready to transmit $\vh^\user_t$. In the encoding step, the model updates are compressed into a single bit using a quantizer and a random binary codebook, and then perturbed to enhance privacy. These steps are formulated as follows.

{\bf Quantization:}
To begin, $\vh^\user_t \in \R^\dimension$ is decomposed into distinct vectors ${\{\updates\}}^M_{\subvec=1}$ such that
\begin{equation}\label{eq:M_definition}
    \updates \in \R^L, \quad 
    M \triangleq 
    \left \lceil{\frac{d}{L}}\right \rceil;
\end{equation}
and being quantized by applying an $L$-dimensional lattice quantizer (Def. \ref{def:lattice_quantizer}) to each, i.e.,  $\updates$ is mapped into $Q_{\cL_t}(\updates)$.

\begin{figure}
\centering
\includegraphics[width=\columnwidth]
{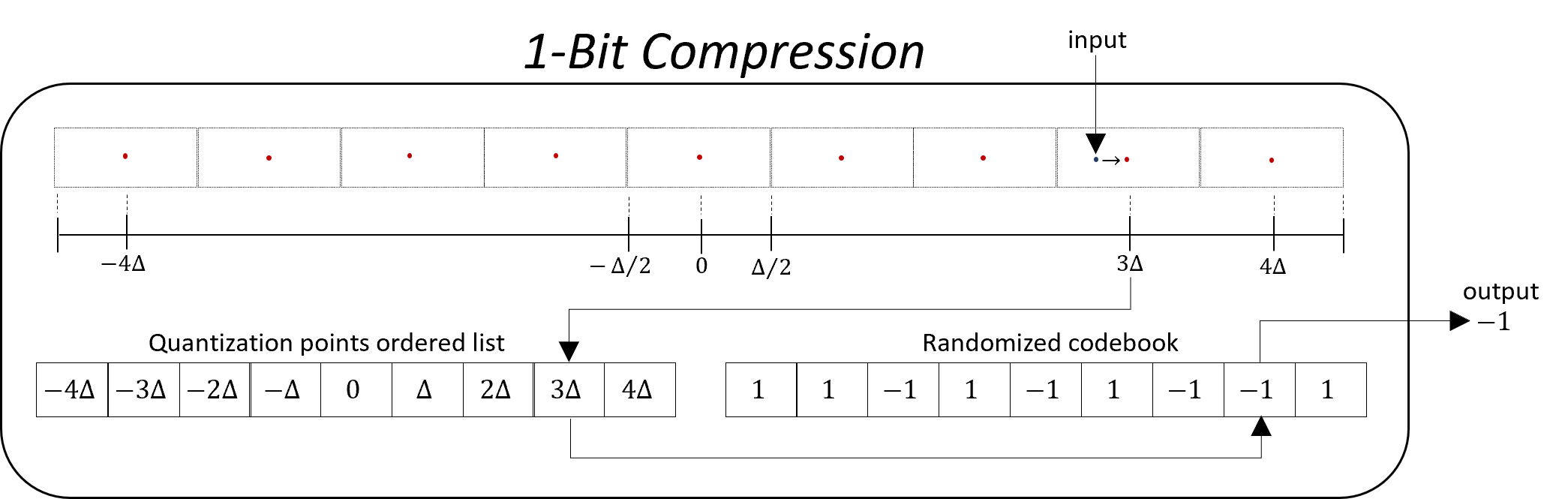}
    \caption{Example: a scalar input is mapped into a point that corresponds to the continues vale of $3\Delta$, which is $9$-th quantization point. Accordingly, as the $9$-th entry of the randomized vector is $-1$, so does the output.}
    \label{fig:toy_example_scalar}
\end{figure}

{\bf 1-Bit Compression:}
Next, the discrete codeword is compressed into a single bit according to the index of the assigned lattice point. To this end, the seed $s_\user$ is used to randomize a codeword $\codeword$, which is uniformly distributed over all words in $\{-1,1\}^{2^{LR}}$ having an equal amount of $1$'s and $-1$'s. 
Then, as illustrated in Fig.~\ref{fig:cpa}, the lattice point $Q_{\cL_t}(\updates)$ is represented by its index in the lattice, denoted $l$, and, in turn, a single bit $\bit$ is set according to the $l$-th entry of the vector $\codeword$. 

Formally, we write $Q_{\cL_t}(\updates)=\Qword$ where $\Qword\in\R^L$ is the $l$-th lattice point in $\cL_t$. The bit that the user conveys to the server is selected based on ${\left[\codeword\right]}_l$, where 
\begin{align}
\label{eqn:Bits}
    \bit\triangleq
    \begin{cases}			
    1 & \text{if } {\left[\codeword\right]}_l=1,\\
    -1 & \text{otherwise}.
    \end{cases}    
\end{align}
The resulting processing converts the $L$ dimensional model updates vector $\updates$ into a single bit representation $\bit$. This procedure is exemplified using a scalar quantizer in Fig.~\ref{fig:toy_example_scalar}.

{\bf Privacy Enhancement:} 
As we show in Section~\ref{sec:analysis}, \kano (Def. \ref{def:kAno}) directly follows from the design of $\codeword$. 
To also maintain $\varepsilon$-\ac{ldp}, \ac{rr} is applied to $\bit$. 
\ac{rr} guarantees $\varepsilon$-\ac{ldp} by having the true value of $\bit$ conveyed with probability $p=\frac{e^\varepsilon}{1+e^\varepsilon}$ and its complement with $1-p$. Consequently, the bit which the sever receives as a representation of  $\updates$ is 
\begin{equation}
\label{eqn:RR}
\LDPbit =
    \begin{cases}
    \bit & {\rm w.p. }\,\, p, \\
   (-1)\cdot \bit & {\rm w.p. }\,\, 1-p.
    \end{cases}
\end{equation}

\subsubsection{Decoding} 
The decoding procedure avoids having the server reconstruct each individual model. Instead, the server uses the bits it receives corresponding to the $\subvec$-th sub-vector of the model parameters to directly compute the desired aggregated model. This is achieved in two stages: first, the bits ${\{\LDPbit\}}_{\user=1}^K$ are recovered into unbiased estimates of their  codewords $\{\codeword\}_{\user=1}^K$, which are directly aggregated into an empirical {\em histogram}, used to update the global model.

{\bf Recovery:} 
Due to the shared seed $s_\user$, the server knows $\codeword$ and can thus associate each bit with its corresponding codeword. However, since the bits are perturbed by the \ac{rr} mechanism, the server can only recover an estimate of the codeword. This is achieved by setting 
\begin{align}\label{eq:reconstructor}
    \RRcodeword=
    \frac{1}{2p-1}
    \begin{cases}
    \codeword& \text{if } \LDPbit=1,\\
    (-1)\cdot\codeword & \text{otherwise};
    \end{cases}
\end{align}
where the weighting factor $\frac{1}{2p-1}$ assures that $\RRcodeword$ is an unbiased estimator of $\codeword$.

{\bf Aggregation:} 
The server then constructs with an aggregated mean of all ${\{\RRcodeword\}}_{\user=1}^K$, i.e., 
\begin{align}\label{eq:RRhistogram_def}
\RRhistogram \triangleq \frac{1}{K}\sum_{\user=1}^K \RRcodeword. 
\end{align}
Practically, $\RRhistogram$ is a discrete normalized histogram. Aggregation through \eqref{eq:RRhistogram_def} can, in principle, yield negative histogram values, which can  be kept or mitigated by thresholding~\cite{thesmar2021cabdriver}. 

The estimated histogram  is utilized for updating the global model, replacing the conventional \ac{fa} update in~\eqref{eq:fed_avg} by 
\begin{align}\label{eq:cpa_update}
    \globalCPA_{t,\subvec} = \globalCPA_{t-\tau,\subvec} + \sum_{l=1}^{2^R} {\left[\RRhistogram\right]}_l \cdot \Qword.
\end{align} 
The global model $\globalCPA_t$ is then obtained by stacking the sub-vectors ${\{\globalCPA_{t,\subvec}\}}_{\subvec=1}^M$. 

\SetKwBlock{Initialization}{Initialization:}{end}
\SetKwBlock{User}{Encode (at the $\user$-th user side, for each $\subvec$):}{end}
\SetKwBlock{Server}{Decode (at the server side, for each $\subvec$):}{end}
\begin{center}
\begin{algorithm}
\caption{1-bit \ac{cpa} at time step $t$}
\label{alg:cpa}
\Initialization{Shared seed $s_\user$, degree of anonymity $k$, privacy budget $\varepsilon$, and lattice $\cL_t$\;}
\User{Quantize $\updates$ into $\Qword$, $l\in \{1,\ldots,2^R\}$, using $Q_{\cL_t}$\;
Set $\bit$ using \eqref{eqn:Bits}\;
Augment $\bit$ into $\LDPbit$ via \eqref{eqn:RR}, convey to  server\;}
\Server{Recover ${\{\RRcodeword\}}_{\user=1}^K$ via \eqref{eq:reconstructor} \;
Obtain an empirical histogram via \eqref{eq:RRhistogram_def}\;
Update the global model $\globalCPA_{t,\subvec}$ using \eqref{eq:cpa_update}\;}
\KwResult{Updated $\subvec$-th global model sub-vector, $\vw_{t,\subvec}$.}
\end{algorithm}
\end{center}

\begin{figure*}
     \centering              
     \includegraphics[width=\textwidth]{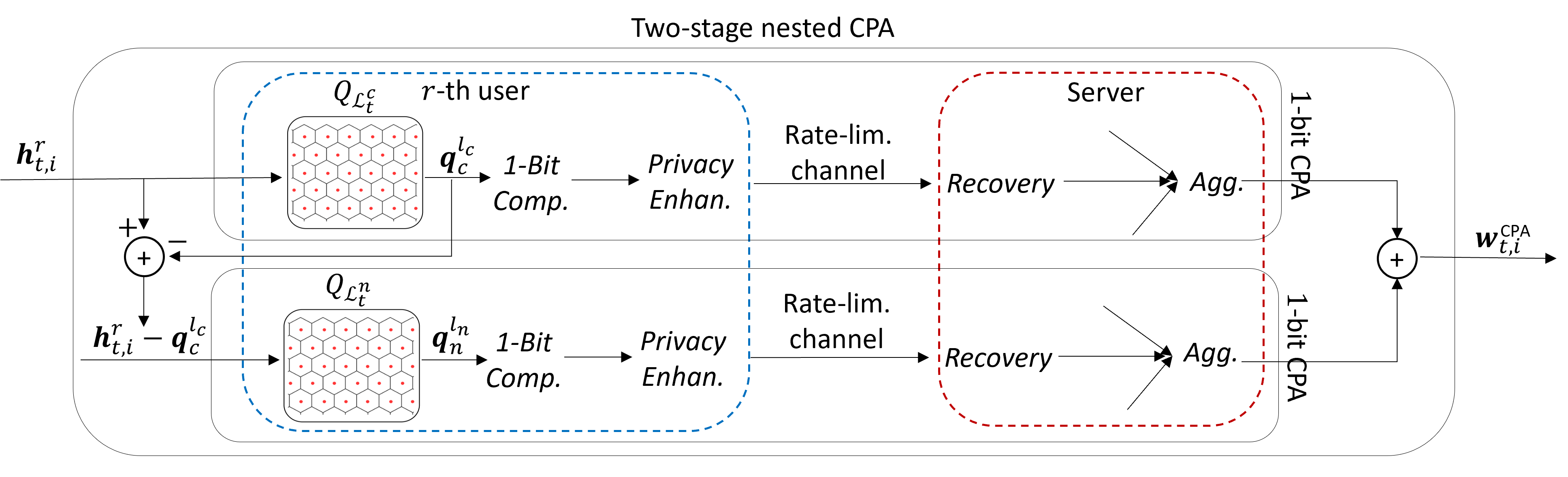}
    \caption{Overview of nested \ac{cpa}. The upper solid box represents the \vcpa with a coarse quantizer while the bottom describes that with the nested one.}
    \label{fig:cpa_with_nq}
\end{figure*}

\subsection{Nested CPA}\label{subsec:nested_cpa}
Algorithm~\ref{alg:cpa} is particularly designed to operate with a large number of users. Instead of recovering the individual model updates sub-vector of each user $\{\updates\}$, the aggregation of $\RRhistogram$ in \eqref{eq:RRhistogram_def} forms an estimate of the {\em distribution} of the quantized updates over the lattice $\cL_t$. As shown in the proof of Theorem~\ref{thm:weights_distortion_bound}, in the horizon of an asymptotically large number of users $K$, the mean value taken over $\RRhistogram$ converges to the federated average of $\{Q_{\cL_t}(\updates)\}_{\user =1}^K$ for any given lattice quantizer employed. This motivates the usage of quantizers with high rate $R$, for which the distortion in $Q_{\cL_t}(\updates)$ compared to $\updates$ is small. However, for a finite number of users, a large number of lattice points typically results in a less accurate estimation of the probability over the lattice via $\RRhistogram$, giving rise to a tradeoff between the number of users $K$ and the quantization rate $R$. 

In order to alleviate this tradeoff, enabling \ac{cpa} to operate reliably with high resolution lattice quantizers (large $R$), we propose to implement fine quantization using multiple low-rate quantizers via {\em nested quantization} (Def.~\ref{def:nestedLQ}). This allows constructing a separate histogram for each low-rate lattice quantizer, such that the mean over the fine lattice, i.e., the desired low-distorted averaged model, can be computed from these histograms. However, this comes at the cost of additional bits conveyed by the users, as each quantized value is no longer conveyed using a single bit as in Algorithm~\ref{alg:cpa}. 
We next formulate this form of nested \ac{cpa}, focusing on a two-stage nested operation (i.e., with two bits $B=\frac{2}{L}$), which can be extended to multiple stages (and multiple bits) as discussed in Subsection~\ref{subsec:quantization}. The general procedure is illustrated in Fig.~\ref{fig:cpa_with_nq}.

\subsubsection{Initialization} 
In addition to the initial steps of Algorithm~\ref{alg:cpa}, nested \ac{cpa} divides the fine lattice $\cL_t$ into a nested lattice $\cL^{\rm n}_t$ and a coarse lattice $\coarse_t$ (see Def.~\ref{def:nestedLQ}), with rates $R_{\rm n}$ and $R_{\rm c}$, respectively. 

\begin{figure}
\centering
\includegraphics[width=\columnwidth]
{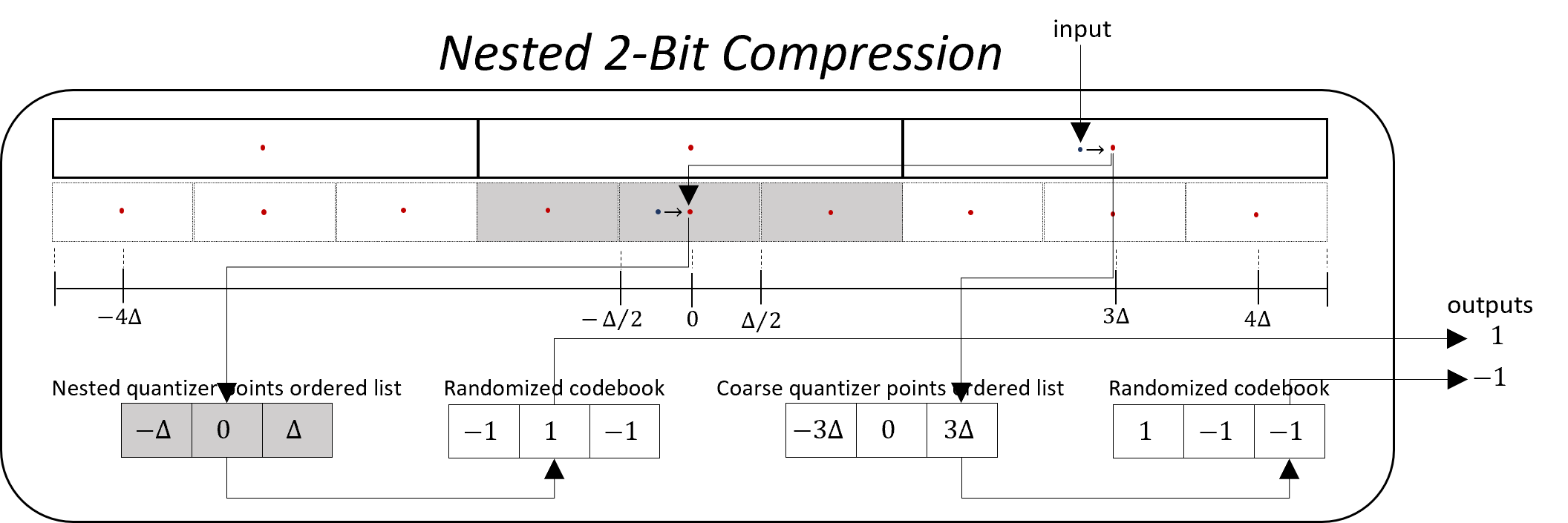}
    \caption{Example: the input is mapped to $3\Delta$ and $0$ by the coarse and nested quantizers respectively; that are the third and second quantization points in each, correspond to $1$ and $-1$ in the randomized vectors, as the outputs.}
    \label{fig:toy_example_nested}
\end{figure}

\subsubsection{Encoding}
As in Algorithm~\ref{alg:cpa}, the model updates are divided into $\{\updates\}_{\subvec=1}^{M}$. Here, the sub-vectors are quantized using the low-rate lattice quantizers, yielding 
\begin{equation}
\label{eqn:LatticeCW}
\begin{split}
    Q_{\coarse_t} ( \updates) = \vq^{l_{\rm c}}_{\rm c},  \\
    Q_{\cL^{\rm n}_t} \big(\updates -   Q_{\coarse_t} ( \updates)\big) =  \vq^{l_{\rm n}}_{\rm n}.
\end{split}
\end{equation}
 The fact that $Q_{\coarse_t}(\cdot)$ and $Q_{\cL^{\rm n}_t}(\cdot)$ form a nested representation of the fine quantizer $Q_{\fine_t}(\cdot)$ indicates that 
 \begin{equation}
\label{eqn:LatticeCWRec}
Q_{\fine_t}(\updates) =    \vq^{l_{\rm c}}_{\rm c} + \vq^{l_{\rm n}}_{\rm n}.
 \end{equation}

Each of the discrete representations in \eqref{eqn:LatticeCW} is then separately mapped into a single bit through the 1-bit compression and privacy enhancement mechanisms detailed in Subsection~\ref{subsec:cpa}. For clarity, the nested extension of the 1-bit compression example given in Fig.~\ref{fig:toy_example_scalar} is depicted in Fig.~\ref{fig:toy_example_nested}.
\subsubsection{Decoding}
Since the server has access to two bits for $\updates$, representing its coarse and nested quantization, it  reconstructs two separate histograms for each lattice. By repeating the steps in \eqref{eq:reconstructor} and \eqref{eq:RRhistogram_def}, the server obtains two histogram estimates: one over $\coarse_t$, denoted  $\RRhistogram^{\rm c} \in \R^{2^{R_{\rm c}}}$, and another over $\cL^{\rm n}_t$ denoted  $\RRhistogram^{\rm n} \in \R^{2^{R_{\rm n}}}$. By \eqref{eqn:LatticeCWRec}, the averaging of $\{  Q_{\fine_t}(\updates)\}_{\user=1}^{K}$ is estimated as the sum of the empirical means over $\coarse_t$ and $\cL^{\rm n}_t$, i.e., the aggregation  in \eqref{eq:cpa_update} is replaced with 
\begin{align}\label{eq:Nestcpa_update}
    \globalCPA_{t,\subvec}\!= \!\globalCPA_{t-\tau,\subvec} \!+\! 
    \sum_{l_c=1}^{2^{R_{\rm c}}} {\left[\RRhistogram^{\rm c} \right]}_{l_{\rm c}} \!\cdot \vq^{l_{\rm c}}_{\rm c} 
    \! +\! 
    \sum_{l_n=1}^{2^{R_{\rm n}}} {\left[\RRhistogram^{\rm n} \right]}_{l_{\rm n}}\! \cdot \vq^{l_{\rm n}}_{\rm n}.
\end{align}

The aggregation rule in \eqref{eq:Nestcpa_update} is designed to approach the corresponding rule obtained using 1-bit \ac{cpa} in Algorithm~\ref{alg:cpa} with the fine lattice $\fine_t$, while using relatively easy-to-estimate histograms over smaller dictionaries. These nested extensions of \ac{cpa} allows to facilitate the learning of an improved model in a federated manner when the number of users $K$ is limited, as numerically evidenced in Section~\ref{sec:experiments}. There, notable improvements of the nested operation over 1-bit \ac{cpa} are observed for $K=10$ users, which vanish for $K=1000$ users. This capability comes at the cost of additional bits being exchanged, though such increased communication is likely to be more tolerable when learning with tens of users compared with learning with thousands and more users.

\subsection{Discussion}\label{subsec:discussion}
The proposed \ac{cpa} is a dual-function mechanism for enhancing privacy while compressing the model updates and aggregating in a robust fashion over large-scale \ac{fl} systems.  It is inspired by the coding scheme of \acl{pmga}~\cite{naim2022private}, with the incorporation of lattice quantization and \ac{ldp} enhancement. The perturbation introduced in \ac{rr}, originating from the need to meet \ref{itm:ldp}, is mitigated here not by the conventional \ac{fa} as in, e.g., \cite{shlezinger2020uveqfed}, but rather by a unique reconstruction of discrete histograms, which in turn guarantees \kano, as we show in Section~\ref{sec:analysis}. 
While \ac{cpa} exploits the high-dimensional structure of the model updates through (possibly high-rate) lattice quantization,
it still dramatically reduces the conventional \ac{fl} communication overhead. This follows since the users transmit at most $B$ bits per sample, and the server avoids recovering the individual updates for each user. The resulting operation is thus scalable and applicable without limiting the number of participants in each \ac{fl} round. In fact, its associated distortion decreases with the number of users, as shown in Section~\ref{sec:analysis}. 

Moreover, because \ac{cpa} inherently limits the influence of a user in the final model, it allows for the training to be Byzantine robust against erroneous adversarial users and poisoning attacks~\cite{yin2018byzantine}, as was also observed for one bit \ac{sgd} with majority vote aggregation in \cite{bernstein2019signsgd} . This is numerically demonstrated in Section~\ref{sec:experiments}. 
We note that \ac{cpa} is expected to enhance privacy also against external adversaries, due to \ac{ldp} post-processing property \cite{xiong2020comprehensive}. However, since \ac{fl} is motivated by the need to avoid sharing local data with a centralized server, characterizing \ac{ldp} guarantees from external adversaries is left for future work. Altogether, \ac{cpa} satisfies \ref{itm:ldp}-\ref{itm:scalability} without notably affecting the utility of the learned model, compared to using separate privacy enhancement and quantization, as numerically demonstrated in Section~\ref{sec:experiments}. 

\ac{cpa} is designed assuming that all users share the same privacy requirement $\varepsilon$ by \ref{itm:ldp}. However, since the encoding  of \ac{cpa} is done separately by each user, one can extend its operation to user-specific privacy budgets. The direct approach simply has all users set their \ac{rr} parameter $p$ to be the lowest non-flipping probability among all utilized \ac{ldp} mechanisms, i.e., the one corresponding to the user with the most strict privacy requirements. Yet, it can possibly be that each user uses a different value of $p$ by modifying the aggregation rule at the server, though this extension and its analysis are left for future work.  
Furthermore, the privacy requirement considered in \ac{cpa} is imposed on each communication round by \ref{itm:ldp}. Traditionally, this requirement can be related to an upper bound on the privacy budget in \ac{fl} over a fixed number of global training rounds by the composition theorem~\cite[Thm. III.1.]{dwork2010boosting}. Alternatively, additional mechanisms can be implemented to avoid accumulating privacy leakage over multiple rounds, as characterized by the composition theorem~\cite{sun2020ldp}. We leave the combination of \ac{cpa} with such mechanisms for future study. 

The nested implementation of \ac{cpa} allows users to convey multiple bits per sample. By doing so, the server is able to construct smaller and therefore more accurate histograms, improving the global model update design in each \ac{fl} round, particularly when $K$ is relatively small. However, with each bit added, the influence of each individual user on the constructed global model, being accordingly updated, grows and increases \ac{cpa}'s sensitivity to malicious users. This gives rise to the existence of a tradeoff between model accuracy, compression, and robustness, whose analysis is left for future work.

\section{Performance Analysis}\label{sec:analysis}
\ac{cpa} is designed to jointly support compression and privacy in \ac{fl} over large-scale networks. Compression directly follows as each user conveys merely $B$ bits per sample. For 1-bit \ac{cpa}, this boils down to merely $M$ bits, i.e., the number of bits is not larger than the number of weights. Consequently, we dedicate this section to theoretically analyze the {\em privacy} and {\em learning} implications of \ac{cpa}, focusing on its 1-bit implementation for simplicity. We characterize its privacy guarantees (Subsection~\ref{subsec:privacy_analysis}), distortion in its recovered global model (Subsection~\ref{subsec:weights_distortion}), and convergence profile (Subsection~\ref{subsec:fl_convergence}). 

\subsection{Privacy Analysis}\label{subsec:privacy_analysis}
In accordance with Requirement \ref{itm:ldp}, we are considering two privacy measures: \ac{ldp} and \kano.
As conventionally done in the private \ac{fl} literature~\cite{wei2020federated, chaudhuri2022privacy, kairouz2021distributed, agarwal2021skellam, lang2022joint}, we characterize the privacy by observing its leakage with respect to the weights for each \ac{fl} round. This stems from the sequential data processing nature of the local learning procedure, which implies that assuring privacy with respect to the model weights guarantees privacy with regard to the datasets.

More specifically, for a given dataset $\cD_{\user}$, each bit produced by \ac{cpa} can be generally viewed as encompassing two subsequent transformations: the first is the training of the model and the quantization of its weights into that corresponding bit, represented by the mapping $f:\cD_{\user} \to \{0,1\}$; and the second is the incorporation of further perturbation via the \ac{rr} mechanism $R(\cdot)$, i.e., $R(f(\cD_{\user}))$. This sequential data-processing form is related to  a conventional result in the literature on differential privacy, stating that $\varepsilon$-\ac{ldp} is not only guaranteed for $f(\cD_{\user})$, but also for $\cD_{\user}$~\cite{dwork2014algorithmic}. This is proven in the following claim:
\begin{claim}
    Let $\cA$ be a finite set, and  $f: \cA \to \{0,1\}$. Let $R: \{0,1\} \to \{0,1\}$ be the \ac{rr} mechanism with differential privacy budget $\varepsilon$. Then, the mechanism $M:\cA \to \{0,1\}$ defined as $M(\vx) = R (f(\vx))$ is an $\varepsilon$-\ac{ldp} mechanism (i.e. any two elements of $\cA$ are $\varepsilon$-indistinguishable).
\end{claim}
\begin{IEEEproof}
    Let $\vx\neq\vy\in \cA$ and $i\in\{0,1\}$. If it holds that $f(\vx)=f(\vy)$, then 
    \begin{align*}
    \proba[M(\vx)=i]=\proba[R\left(f(\vx)\right)=i]=
    \proba[R\left(f(\vy)\right)=i]\\=
    \proba[M(\vy)=i].
    \end{align*}
    That is, since $\proba[M(\vx)=i]=\proba[M(\vy)=i]$; we actually get that $\vx,\vy$ are $0$-indistinguishable. 
    Otherwise, the two bits satisfy $f(\vx)\neq f(\vy)$, and since $R(\cdot)$ is the \ac{rr} mechanism with differential privacy budget $\varepsilon$, it satisfies
    \begin{align*}
    \proba[M(\vx)=i]=\proba[R\left(f(\vx)\right)=i]\leq e^{\varepsilon}\proba[R\left(f(\vy)\right)=i]\\
    =e^{\varepsilon}\proba[M(\vy)=i].\end{align*}
    Thus, $\proba[M(\vx)=i]\leq e^{\varepsilon} \proba[M(\vy)=i]$, and $M$ is an $\varepsilon$  locally differentially private mechanism on $\cA$.
\end{IEEEproof}
We note that the setting where  $f(\vx)=f(\vy)$, i.e., two different datasets lead to having an identical mapping (due to quantization), is actually where the \kano of \ac{cpa} appears. 

The encoding steps of \ac{cpa} are directly  derived to meet the definitions of both privacy measures, \ac{ldp} and \kano, as formally stated in the following propositions.

\begin{proposition}\label{pro:LDP}
\ac{cpa} is $\varepsilon$-\ac{ldp} with respect to $\cD_\user$, per communication round.
\end{proposition}
The \ac{ldp} guarantee in Proposition~\ref{pro:LDP} follows solely from the usage of \ac{rr}, and is invariant of the preceding processing of \ac{cpa}. 
While in this work we further utilize probabilistic quantizers only for \ac{cpa}'s distortion analysis (Subsection~\ref{subsec:weights_distortion}), one can possibly additionally enhance privacy in the algorithm's quantization stage by a proper design of this stochastic compression technique, as shown in \cite{lang2022joint}, though this extension is left for future study. Nevertheless, the quantization stage plays a key role in achieving \kano, as stated next.

\begin{proposition}\label{pro:kano}
\ac{cpa} preserves \kano with respect to the lattice quantization of $\updates$. 
\end{proposition}
\begin{IEEEproof}
\kano (Def. \ref{def:kAno}) follows by-construction from \ac{cpa}'s design in Algorithm~\ref{alg:cpa}. There, a user's update $\vh^r_t$ is decomposed into sub-vectors $\{\updates\}_{\subvec=1}^{M}$. 
Each $\updates\in \R^L$ is quantized into one codeword out of all possible codewords, say of index $l$, $\vq_l=Q_{\cL_t}(\updates)$; which is in turn mapped into a binary vector $\codeword$ of length equals to the total number of codewords. $\codeword$ is comprised of equal amount of $1$'s and $-1$'s in expectation and assured to have $1$ in its $l$th entry (see Fig.~\ref{fig:cpa}).

Therefore, as half of $\codeword$'s entries are expected to be $1$, it is implied that $k$ values - being half of codewords - are valid candidates mappings of $\updates$. 
Since a vector quantizer of dimension $L$ and rate $R$ has $2^{RL}$ codewords, $1$-bit \ac{cpa} with such a quantizer has the server not able to distinguish between expected $k\triangleq 2^{LR}/2=2^{LR-1}$ adequate values.
\end{IEEEproof}
\smallskip

The \kano of \ac{cpa} guarantees the indistinguishability between $k$ possible values of the lattice quantization of $\updates$, where $k=2^{LR-1}$ is solely determined by the quantization dimension and rate parameters. Both can be tuned to fix a desirable value of $k$ in parallel to trade-offing privacy and compression considerations.

While Proposition~\ref{pro:kano} formulates the anonymity degree of each sub-vector, Corollary~\ref{cor:kano} reveals the higher degree of anonymity achieved with respect to the complete model.
\begin{corollary}\label{cor:kano}
\ac{cpa} preserves $k^M$ anonymity with respect to the lattice quantization of $\vh^\user_t$.
\end{corollary}
\begin{IEEEproof}
The corollary follows directly from Proposition~\ref{pro:kano}. The server cannot distinguish between $k$ different possibilities for each $\updates$ and $\vh^\user_t$ is a concatenation of $\{\updates\}_{\subvec=1}^{M}$. Thus, each set of bits can represent $k^M$ different $\vh^\user_t$ settings.
\end{IEEEproof}

Proposition~\ref{pro:kano} in fact follows from the nature of the \kano measure, which is defined over finite sets (see Def.~\ref{def:kAno}). However, by the operation of the quantization procedure, it collapses all the continuous values within a certain cell into the same decision, i.e., lattice point.  This can be interpreted as a kind of \say{continuous} \kano, as, in addition to the indistinguishability between $k$ different lattice points guaranteed by Proposition~\ref{pro:kano}, a quantized private value can potentially originate from any item in the uncountable set that covers the decision area mapped into this lattice point.

\subsection{Weights Distortion}\label{subsec:weights_distortion} 
\ac{cpa} incorporates lossy compression,  \ac{rr}, and a unique aggregation formulation that deviates from the conventional \ac{fa}. All of these steps inherently induce some distortion on the model updates, compared to the desired average of the model updates. 
To quantify this distortion, we characterize the difference between the model aggregated via \ac{cpa} denoted $\globalCPA_{t+\tau}$ obtained by stacking ${\{\globalCPA_{t,\subvec}\}}_{\subvec=1}^M$  in~\eqref{eq:cpa_update}, with the desired average  (whose direct computation gives rise to communication and privacy considerations) obtained from the same global model at time $t$, given by
\begin{align}\label{eq:globalFA}
    \globalFA_{t+\tau} \triangleq \globalCPA_t + \frac{1}{K}\sum_{\user=1}^K \vh^\user_{t+\tau}.
\end{align}

Next, we show that the effect of the excessive distortion induced by \ac{cpa} can be mitigated while recovering the desired $\globalFA_{t+\tau}$ as $\globalCPA_{t+\tau}$. Thus, the accuracy of the global learned model is maintained despite the incorporation of the scalable compression and privacy mechanisms of \ac{cpa}.
In our analysis we adopt the following assumption
\begin{enumerate}[label={\em AS\arabic*},series=assumptions]     
    \item \label{itm:DQ}
    \ac{cpa} uses {\em probabilistic quantization}, i.e., $Q_{\cL}(\cdot)$ implements {\em \ac{dq}}.
\end{enumerate}
Probabilistic quantizers, as assumed in \ref{itm:DQ}, are often employed in \ac{fl}, due to the stochastic nature of their associated distortion~\cite{konevcny2016federated,shlezinger2020uveqfed,alistarh2017qsgd,reisizadeh2020fedpaq}, \cite{lang2022joint}. 
For a given probabilistic lattice quantizer defined using the lattice $\cL$ with lattice points $\{\Qword\}_{l=1}^{2^R}$, 
we define $\latticeM$ to be the normalized second-order moment of this lattice, or alternatively, the variance of the distortion induced by \ac{dq} with lattice $\cL$, as
\begin{equation}\label{eq:normalizedsecond-order moment of the lattice}
\latticeM \triangleq \frac{\frac{1}{L}\int_{\cP_0} \|\vx\|^2d\vx}{\text{vol}(\cP_0)^{2/L}};      
\end{equation}
where $L$ is the lattice dimension, `vol' stands for `volume', and $\cP_0=\{\vx:Q_\cL(\vx)=\boldsymbol 0\}$ is the basic lattice cell (see further details in Section~\ref{subsec:quantization}).
For \ac{cpa} employing this lattice quantizer, the distance between the recovered model $\globalCPA_{t+\tau}$ and the desired one $\globalFA_{t+\tau}$ satisfies:
\begin{theorem}\label{thm:weights_distortion_bound}
When Assumption~\ref{itm:DQ} holds, the mean-squared distance between $\globalCPA_{t+\tau}$ and $\globalFA_{t+\tau}$ is bounded by  
\begin{align}\label{eq:weights_distortion_bound}
    \E\left[{\big\|\globalCPA_{t+\tau}\!-\!\globalFA_{t+\tau}\big\|}^2\right]\leq
       \frac{M}{K}\left(\sum_{l=1}^{2^R}\frac{\left\|\Qword\right\|^2}{{\left(2p\!-\!1\right)}^2}
     \!+\!\latticeM\right).
\end{align}
\end{theorem}
\numberwithin{lemma}{subsection} 
\numberwithin{corollary}{subsection} 
\numberwithin{remark}{subsection} 
\numberwithin{equation}{subsection}	
\begin{IEEEproof}
To prove Theorem~\ref{thm:weights_distortion_bound}, we separate the  distortion induced by \ac{cpa}   into two terms and bound each of them separately. The first term is the error caused by the discrete histogram estimation  in \eqref{eq:RRhistogram_def};  the latter is the compression distortion.

As in Subsection \ref{subsubsec:encoding},  we denote by ${\{\vv_\subvec\}}_{\subvec=1}^M$ the decomposition of a vector $\vv$ into $M$ distinct $L\times 1$ sub-vectors. Moreover, we write $\{c^l\}$ as the quantizer  words' rates, given by $c^l\triangleq\frac{1}{K}\sum_{\user=1}^K \delta\left[Q_{\cL_t}(\updatesTau)-\Qword\right]$, where $\delta[\cdot]$ is the Kronecker delta.
By \eqref{eq:globalFA} and \eqref{eq:cpa_update} we have that
\begin{align}
&\E\left[{\|\globalCPA_{t\!+\!\tau, \subvec}\!-\!\globalFA_{t\!+\!\tau, \subvec}\|}^2\right]
\!=\!
\E\left[{\left\|\frac{1}{K}\sum_{\user=1}^K \updatesTau \!-\! \sum_{l=1}^{2^R} {\left[\RRhistogramTau\right]}_l\Qword
\right\|}^2\right] \notag \\
&\quad \stackrel{(a)}{=} \E\bigg[\bigg\|\Big(\frac{1}{K}\sum_{r=1}^K \updatesTau - Q_{\cL_t}(\updatesTau)\Big) \notag \\
& \qquad \qquad - \Big(\sum_{l=1}^{2^R} \left(c^l - {\left[\RRhistogramTau\right]}_l\right)\Qword\Big)\bigg\|^2\bigg],
\label{eq:before_defining_cl}
\end{align}
where $(a)$ is obtained by adding and subtracting $\frac{1}{K}\sum_{r=1}^KQ_\cL\left(\updatesTau\right) = \sum_{l=1}^{2^R} c^l\cdot\Qword$.

To proceed, we define the compression distortion $\QerrorTau\triangleq Q_{\cL_t}(\updatesTau) -\updatesTau$ and the histogram estimation error $\HerrorTau\triangleq c^l - {\left[\RRhistogramTau\right]}_l$. The joint distribution of $\{\QerrorTau\}, \{\HerrorTau\}$ is characterized in the following lemma:
\begin{lemma}\label{lemma:DQ_distortion}
For any $\{\updatesTau\}$ it holds that the sequences $\{\QerrorTau\}$ and $\{\HerrorTau\}$ are uncorrelated (over $r$ and $l$, receptively), zero-mean, and mutually uncorrelated. The variance of $\QerrorTau$ equals   $\latticeM$ while that of $\HerrorTau$ is  bounded by $\frac{1}{K\cdot{(2p-1)}^2}$.
\end{lemma}
\begin{IEEEproof}
By Assumption~\ref{itm:DQ}, $Q_{\cL_t}$ realizes \ac{dq}, and thus its distortion is zero-mean i.i.d. with variance $\latticeM$ \cite{gray1993dithered,zamir1996lattice, shlezinger2019hardware}. Combining this with the independence of the dither, the codewords, and the \ac{rr}  implies that $\QerrorTau$ and  $\HerrorTau$ are uncorrelated. 
 
For $\Herror$, we define $\tilde{\eta}_{t,i}^{l,r}=\delta[Q_\cL(\vh^r_{t,i})-\vq_l] - \left[\tilde\vv^r_{t,i}\right]_l$. By the definitions of the codeword and  \ac{rr}, it holds that for any $\{\vh^r_{t,i}\}$, 
    \begin{align*}
    \tilde{\eta}_{t,i}^{l,r}&=  
    \begin{cases}
        \begin{cases}
        1 - \frac{1}{2p-1},\text{ w.p } p\\
        1 + \frac{1}{2p-1},\text{ w.p } 1-p
        \end{cases}    &\text{ if } \delta[Q_\cL(\vh^r_{t,i})-\vq_l] =1,\\
        \pm \frac{1}{2p-1} \text{ w.p } 0.5 &\text{otherwise};
    \end{cases}
    \end{align*}  
    and thus $\HerrorTau=\frac{1}{K}\sum_{r=1}^K  \tilde{\eta}_{t+\tau,i}^{l,r}$ are i.i.d. zero-mean with variance not larger than $\frac{1}{K\cdot{(2p-1)}^2}$.
\end{IEEEproof}
Altogether, combining \eqref{eq:before_defining_cl} with Lemma~\ref{lemma:DQ_distortion} yields
\begin{align*}
&\E\left[{\|\globalCPA_{t+\tau}-\globalFA_{t+\tau}\|}^2\right]
=\sum_{\subvec=1}^M\E\left[{\|\globalCPA_{t+\tau, \subvec}-\globalFA_{t+\tau, \subvec}\|}^2\right]\\
&= \sum_{\subvec=1}^M \E\bigg[\Big\|\frac{1}{K}\sum_{r=1}^K \QerrorTau\Big\|^2\bigg]  + 
\E\bigg[\Big\|\sum_{l=1}^{2^R} \HerrorTau \Qword \Big\|^2\bigg] \notag \\
 &\leq\frac{M}{K}\latticeM  + \frac{M}{K{(2p-1)}^2}\sum_{l=1}^{2^R} {\left\|\Qword\right\|}^2,
\end{align*}
thus proving \eqref{eq:weights_distortion_bound}.  
\end{IEEEproof}
\smallskip
  
Theorem~\ref{thm:weights_distortion_bound} implies that the recovered model can be made arbitrarily close to the desired one by increasing the number of edge users participating in the \ac{fl} training procedure, as \eqref{eq:weights_distortion_bound} decreases as $1/K$. This holds as there, 
the histogram estimation error term, i.e., $\frac{1}{K}\sum_{l=1}^{2^R}\frac{\left\|\Qword\right\|^2}{{\left(2p\!-\!1\right)}^2}$ 
accounts for the distance between the histogram's empirical evaluation to its true value; which is the distance between the empirical mean of the i.i.d-randomized codewords $\{\codeword\}_{\user=1}^K$ and its expected value. By the law of large numbers, this distance approaches zero with growing $K$ while its associated variance decreases at a rate of $1/K$.

The exact same arguments also accounts for the compression distortion term $\frac{1}{K}\latticeM$; due to the unique framework of \ac{fl}, where the users quantized quantities are only taken in {\em average}. In \ac{cpa}, the employed compression is a probabilistic one, for which the output can be modeled as the input plus an i.i.d additive-noise of mean zero and bounded variance.  
As a result, adding more users does not induce more quantization errors, but is actually a contributing factor that improves the empirical estimations of \ac{cpa}. 
This indicates the suitability of \ac{cpa} for \ac{fl} over large networks. 

However, while the difference decaying rate is of order $\cO(1/K)$, it is still affected by the need to compress the model updates and enhance their privacy. This is revealed in Theorem~\ref{thm:weights_distortion_bound} via the presence of $M,\latticeM$ and $p$, arising from each consideration. In particular, $M \triangleq \ceil{\frac{d}{L}}$ stands for the number of distinct $L \times 1$ sub-vectors in $\vh^\user_t \in \R^\dimension$, i.e., ${\{\updates\}}^M_{\subvec=1}$, where each is being quantized by applying an $L$-dimensional lattice quantizer. Consequently, lower $M$, or equivalently, higher quantization dimension $(L)$, particularly (physically) implies that more entries of $\vh^\user_t$ are being quantized together. In vector quantitation theory \cite[Part V]{polyanskiy2014lecture}, this is known to improve compression performance; what further explains why \eqref{eq:weights_distortion_bound} linearly depends on $M$. 
The moment $\latticeM$ follows from the distortion induced by lattice quantization, while $p$ accounts for the distortion induced by the \ac{rr} mechanism. Specifically, $p$ is dictated by the privacy level $\varepsilon$, which implies that, as expected, stricter privacy constraints lead to additional distortion in the recovered global model.

\subsection{Federated Learning Convergence}\label{subsec:fl_convergence}
In the previous subsection we bounded the distortion induced by \ac{cpa} in each communication round to achieve privacy and compression. Next, we show that this property is translated into \ac{fl} convergence guarantees. To that aim, we further introduce the following assumptions, that are commonly employed in \ac{fl} convergence studies in, e.g., \cite{shlezinger2020uveqfed,li2019convergence, stich2018local}, on the local datasets, stochastic gradients, and objectives:
\begin{enumerate}[resume*=assumptions]
    \item \label{itm:heterogenity}
    Each dataset $\cD_\user$ is comprised of i.i.d samples. However, different datasets can be statistically heterogeneous, i.e., arise from different distributions. 
    \item \label{itm:bounded_norm}
    The expected squared $\ell_2$-norm of the vector $\nabla F^{\sample}_\user\left(\vw\right)$ in \eqref{eq:local_sgd} is bounded by some $\xi^2_\user > 0$ for all $\vw\in\R^\dimension$.
    \item \label{itm:obj_smooth_convex}
    The  objective functions $\{F_\user(\cdot)\}^K_{\user=1}$ are  $\rho_s$-smooth and $\rho_c$-strongly convex, i.e., for all $\vw_1, \vw_2 \in \R^\dimension$ we have
    \begin{multline*}
     (\vw_1 -\vw_2)^T\nabla F_\user(\vw_2)
        +\frac{1}{2}\rho_c{\|\vw_1 -\vw_2\|}^2 \\
      \leq    F_\user(\vw_1)-F_\user(\vw_2)  \leq \\
        (\vw_1 -\vw_2)^T\nabla F_\user(\vw_2)
        +\frac{1}{2}\rho_s{\|\vw_1 -\vw_2\|}^2.
    \end{multline*}
\end{enumerate}

Statistical heterogeneity as in \ref{itm:heterogenity} is a common characteristic of \ac{fl} \cite{kairouz2021advances,li2020federated,gafni2021federated}. It is consistent with Requirement \ref{itm:universal}, which does not impose any specific distribution on the data. Statistical heterogeneity implies that the local objectives differ between users, hence the dependence on $\user$ in \ref{itm:bounded_norm}, often employed in distributed learning studies \cite{shlezinger2020uveqfed,li2019convergence, stich2018local}. 
Following \ref{itm:heterogenity} and \cite{li2019convergence, shlezinger2020uveqfed, lang2022joint, lang2024stragglers}, we define the heterogeneity gap,
\begin{align}\label{eq:psi_heterogeneity_gap}
\psi \triangleq F(\vw^{\rm opt})-\frac{1}{K}\sum_{\user=1}^K \min_{\vw} F_\user(\vw),
\end{align}
where $\vw^{\rm opt}$ is defined in \eqref{eq:w_opt}, and $F(\vw^{\rm opt})$, $\min_{\vw} F_\user(\vw)$ are the minimum values of $F, \ F_\user$, respectively. Eq. \eqref{eq:psi_heterogeneity_gap} quantifies the degree of non-i.i.d: for i.i.d. data, $\psi$ approaches zero as the number of samples grows; and otherwise, non-i.i.d. data results with nonzero $\psi$, and its magnitude reflects the heterogeneity of the data distribution.
Finally, assumption \ref{itm:obj_smooth_convex} holds for a range of objective functions used in \ac{fl}, including $\ell_2$-norm regularized linear regression and logistic regression \cite{shlezinger2020uveqfed}. 

It is emphasized, though, that assumptions \ref{itm:DQ}-\ref{itm:obj_smooth_convex} are only introduced for having a tractable analysis, and we further empirically demonstrate the usefulness of \ac{cpa}, whose derivation is invariant to these assumptions, in settings where they do not necessarily hold as exemplified in Section~\ref{sec:experiments}.

The following theorem characterizes the convergence of \ac{fl} employing \ac{cpa} with \lsgd training:
\begin{theorem}\label{thm:FL_Convergence}
Let $\cL$ be a lattice with generator matrix $\vG$, moment $\latticeM$, and points $\{\Qword\}_{l=1}^{2^R}$. Set $\varphi \triangleq \tau \max \left(1 , 4\rho_s / \rho_c \right)$. Then consider \ac{cpa}-aided \ac{fl} satisfying 
\ref{itm:DQ}-\ref{itm:obj_smooth_convex} while using, at each round $t$, a lattice quantizer $\cL_t$ with generator matrix $\vG_t=\zeta_t\cdot \vG$, where $\zeta_t$ is a positive sequence holding $ \zeta_t^2 \leq C\cdot \eta_t^2$ for some fixed $C>0$. Under this setting, \lsgd with step-size $\eta_t=\frac{\tau}{\rho_c(t+\varphi)}$ for each $t\in\N$ satisfies
\begin{align}\label{eq:FL_Convergence}
    \notag&\E\left[F(\globalCPA_t)\right]-F(\vw^{\rm opt})\leq\\
    &\qquad \frac{\rho_s}{2(t+\varphi)}
    \max\left(\frac{\rho^2_c+\tau^2 \ConvCoeff}{\tau\rho_c^2},
    \varphi\|\vw_0 - \vw^{\rm opt}\|^2\right),
\end{align}
where
\begin{align}
\ConvCoeff \triangleq
     \frac{1}{K}
    \Bigg\{&
    M\cdot C\cdot\Big(
    \sum_{l=1}^{2^R}\frac{{\left\|\vq^l\right\|}^2}{{\left(2p-1\right)}^2}+\latticeM\Big) 
   + \frac{1}{K}\sum_{r=1}^K\xi^2_r \notag \\
   &+ 8{\left(\tau-1\right)}^2\sum_{r=1}^K\xi^2_r
    \Bigg\} + 6\rho_s\psi.
    \label{eq:FL_convergence_b_term} 
\end{align}
\end{theorem}
\numberwithin{lemma}{subsection} 
\numberwithin{corollary}{subsection} 
\numberwithin{remark}{subsection} 
\numberwithin{equation}{subsection}	
\begin{IEEEproof}   
To prove the theorem, we derive a recursive bound on the weights error, from which the \ac{fl} convergence bound is then concluded. This outline follows the steps used in \cite{shlezinger2020uveqfed}, which did not consider privacy or anonymity guarantees. 
We next briefly describe the  main steps for completeness,  deferring the  proofs of some of the intermediate lemmas to \cite{shlezinger2020uveqfed}.
\subsubsection*{Recursive Bound on Weights Error}
Denote by $\cI_\tau\triangleq\{n\tau|n=1,2, \dots\}$ the set of global synchronization steps, i.e., $t+1 \in \cI_\tau$ is a time step indicates communication of all devices. 
Aiding these notations, \ac{cpa} induces excessive distortion (compared to vanilla \ac{fa}) in each time instance in $\cI_\tau$. This can be formally written as
\begin{align}
    &\notag\weights_{t+1}=
    \begin{cases}
        \weights_t-\eta_t \nabla F^{\sample}_\user\left(\weights_t\right) & t+1\not\in\cI_\tau,\\
        \globalCPA_t  & t+1\in\cI_\tau;\\
    \end{cases}    \\
    \overset{(a)}{\triangleq}
    &\begin{cases}
        \weights_t-\eta_t \nabla F^{\sample}_\user\left(\weights_t\right)+ \underbrace{\CPAerror}_{=0} & t+1\not\in\cI_\tau,\\
        \frac{1}{K}\sum\limits_{\user'=1}^K \left(
       \weightsTag_t-\eta_t \nabla F^{\sampleTag}_{\user'}\!\big(\weightsTag_t\big)
        \!+\!\CPAerror\right) & t+1\in\cI_\tau;
    \end{cases}\label{eq:w_t_r_def}
\end{align}
where $(a)$ follows from the subtraction and addition of $\globalFA_t$ defined in \eqref{eq:globalFA}, and setting $\CPAerror\triangleq \globalCPA_t-\globalFA_t$.

We next define a virtual sequence $\{\virtual_t\}$ from $\{\weights_t\}$
that coincides with $\globalCPA_t$ for $t \in \cI_\tau$. Specifically,
\begin{align}
    \notag\virtual_{t+1} 
    &\triangleq 
    \frac{1}{K}\sum_{\user=1}^K 
    \weights_{t+1}
    \overset{\eqref{eq:w_t_r_def}}{=}
    \frac{1}{K}\sum_{\user=1}^K 
    \left(\weights_t-\eta_t \nabla F^{\sample}_\user\left(\weights_t\right)+ \CPAerror \right)\\
    &= \virtual_t
    - \eta_t\underbrace{\frac{1}{K}\sum_{\user=1}^K 
    \left(\nabla F^{\sample}_\user\left(\weights_t\right)
    -\frac{1}{\eta_t}\CPAerror \right)}_{\triangleq \Noisygrad}. 
    \label{eq:noisy_grad}
\end{align}
In \eqref{eq:noisy_grad}, $\Noisygrad$ is the averaged noisy stochastic gradient, where  the averaged full gradient are
    $\grad \triangleq \frac{1}{K}\sum_{\user=1}^K 
    \nabla F_\user\left(\weights_t\right)$.

The resulting model is thus equivalent to that used in \cite[App. C]{shlezinger2020uveqfed}. Thus, by  \ref{itm:obj_smooth_convex} it follows that if $\eta_t\leq\frac{1}{4\rho_s}$ then
\begin{multline}\label{eq:expected_distance_bound}
 \!   \E\left[\left\| 
    \virtual_{t+1}\!-\!\vw^{\rm opt} \right\|^2\right]    \leq 
    (1\!-\!\eta_t\rho_c) 
    \E\left[\left\| \virtual_t\!-\!\vw^{\rm opt} \right\|^2\right]    +  6\rho_s\eta^2_t\psi\\
     +
     \eta^2_t \E\left[\left\| \Noisygrad-\grad \right\|^2\right]   + 2\E\left[\frac{1}{K}\sum_{\user=1}^K
     \left\|\virtual_t-\weights_t\right\|^2\right].
\end{multline} 
Equation \eqref{eq:expected_distance_bound} bounds the expected distance between the virtual sequence $\{\virtual_t\}$ and the optimal weights $\vw^{\rm opt}$ in a recursive manner. We further bound the summands in \eqref{eq:expected_distance_bound}:

\begin{lemma}\label{lemma:g_functions_bound}
If the step-size $\eta_t$ is non-increasing and satisfies $\eta_t \leq 2\eta_{t+\tau}$ for each $t\geq 0$, then, when  \ref{itm:bounded_norm} holds,  we have
 \begin{align}\label{eq:g_functions_bound}   
 \notag&\eta^2_t \E\left[\left\| \Noisygrad-\grad \right\|^2\right] \leq \\
&\frac{\eta^2_t}{K}
 \left(
    M\cdot C\left(    \sum_{l=1}^{2^R}\frac{{\left\|\vq^l\right\|}^2}{{\left(2p-1\right)}^2}+\latticeM\right) 
   +\frac{1}{K} \sum_{r=1}^K\xi^2_r    
    \right). 
\end{align}
\end{lemma}
\begin{IEEEproof}\label{proof:g_functions_bound_proof}
To prove \eqref{eq:g_functions_bound}, we separate it into two independent terms and bound each separately. Specifically, 
\begin{align}
    &\eta^2_t\E\left[{
    \left\|\Noisygrad - \grad\right\|}^2\right] =
    \E\left[{\left\|\CPAerror 
    \right\|}^2\right]  \label{eq:cpa_term} \\
    &+\eta^2_t\E\left[{\left\| \frac{1}{K    }\sum_{\user=1}^K
    \left(\nabla F^{\sample}_\user\left(\weights_t\right) - \nabla F_\user\left(\weights_t\right)\right)\right\|}^2\right]  \label{eq:gradients_term}\\
    &-\frac{\eta_t}{K}\sum_{\user=1}^K\E\left[ \CPAerror^T\left(\nabla F^{\sample}_\user\left(\weights_t\right) - \nabla F_\user\left(\weights_t\right)\right)    \right].\label{eq:crossing_cpa_gradients}
\end{align}
We first note that  $\text{\eqref{eq:crossing_cpa_gradients}}=0$ as the stochastic gradients are unbiased estimates of the true gradients and are independent of the distortion. For the first term, it holds that
\begin{align}
    \notag\eqref{eq:cpa_term} 
    \overset{(a)}{\leq} 
     &\frac{M}{K}\left(\sum_{l=1}^{2^R}\frac{\left\|\Qword\right\|^2}{{\left(2p-1\right)}^2}
     +\latticeM\right) \notag \\
     &\overset{(b)}{\leq} 
     \frac{\eta_t^2M\cdot C}{K}\left(\sum_{l=1}^{2^R}\frac{\left\|\Qword\right\|^2}{{\left(2p-1\right)}^2}+\latticeM\right),\label{eq:cpa_term_bound}
\end{align}
where $(a)$ follows by Theorem~\ref{thm:weights_distortion_bound}; $(b)$ holds as \ac{cpa} here realizes a time-variant lattice quantizer $\cL_t$ with a bounded scaled generator matrix. 
For the second term we have
\begin{align}    \notag\eqref{eq:gradients_term}&\overset{(a)}{=}   \frac{\eta^2_t}{K^2}\sum_{\user=1}^K
    \E\left[{\left\| 
    \nabla F^{\sample}_\user\left(\weights_t\right) - \nabla F_\user\left(\weights_t\right)\right\|}^2\right] \notag \\
    &\overset{(b)}{\leq}
    \frac{\eta^2_t}{K^2}\sum_{\user=1}^K \xi^2_\user,\label{eq:gradients_term_bound}
\end{align}
where (a) follows form the uniform distribution of 
the random index $\sample$; and $(b)$ stems from~\ref{itm:bounded_norm}.
Combining  \eqref{eq:cpa_term_bound} and \eqref{eq:gradients_term_bound} concludes the proof.
\end{IEEEproof}

\begin{lemma}\label{lemma:v_w_functions_bound}
If the step-size $\eta_t$ is non-increasing and satisfies $\eta_t \leq 2\eta_{t+\tau}$ for each $t\geq 0$, then when  \ref{itm:bounded_norm} holds, we have
\begin{align}\label{eq:v_w_functions_bound}
    \E\left[\frac{1}{K}\sum_{k=1}^K
    \left\|\virtual_t-\weights_t \right\|^2\right] \leq 
    \frac{4(\tau-1)^2
    \eta_{t}^2}{K}\sum_{\user=1}^K \xi^2_\user,
\end{align}
\end{lemma}
\begin{IEEEproof}
    The proof of Lemma \ref{lemma:v_w_functions_bound} is given in~\cite[App. C]{shlezinger2020uveqfed}. 
\end{IEEEproof}

Next, we define $\delta_t\triangleq
\E\big[
{\left\| \virtual_t-\vw^{\rm opt} \right\|}^2
\big]$, which represents the $\ell_2$-norm of the error in the weights of the global model for $t\in\cI_\tau$. Using Lemmas \ref{lemma:g_functions_bound}-\ref{lemma:v_w_functions_bound}, while integrating \eqref{eq:g_functions_bound} and \eqref{eq:v_w_functions_bound} into \eqref{eq:expected_distance_bound}, we obtain the following recursive relationship:
\begin{align}\label{eq:recursive_delta}
    \delta_{t+1}&\leq(1-\eta_t\rho_c)\delta_t+\eta^2_tb,    
\end{align}
where $\ConvCoeff$ is defined in \eqref{eq:FL_convergence_b_term}.
The relationship in \eqref{eq:recursive_delta} is used in the sequel to prove the \ac{fl} convergence bound in \eqref{eq:FL_Convergence}.

\subsubsection*{FL Convergence Bound}
We next obtain the   convergence bound  by  setting the step-size and the \ac{fl}  parameters in \eqref{eq:recursive_delta} to bound $\delta_t$; and combine the resulting bound with \ref{itm:obj_smooth_convex} to prove \eqref{eq:FL_Convergence}.
In particular, we set  $\eta_t$ to take the form $\eta_t=\frac{\beta}{t+\varphi}$ for some $\beta>0$ and $\varphi \geq \max (4\rho_s \beta, \tau)$, for which $\eta_t\leq\frac{1}{4\rho_s}$ and $\eta_t\leq 2 \eta_{t+\tau}$, implying that \eqref{eq:expected_distance_bound}, \eqref{eq:g_functions_bound}, and \eqref{eq:v_w_functions_bound} hold. 
Under such settings, in  \cite[App. C]{shlezinger2020uveqfed} is it proved  that for $\lambda\geq\max\left(\frac{1+\beta^2b}{\beta\rho_c},\varphi\delta_0\right)$ is holds that $\delta_t\leq\frac{\lambda}{t+\varphi}$ for all integer $t\geq0$.
Finally, the smoothness of the objective~\ref{itm:obj_smooth_convex} implies that
\begin{align}\label{eq:smoothness_implies}
    \E\left[F(\globalCPA_t)\right]-F(\vw^{\rm opt})\leq\frac{\rho_s}{2}\delta_t\leq
    \frac{\rho_s\lambda}{2(t+\varphi)}.
\end{align}
Setting $\beta=\frac{\tau}{\rho_c}$ results in $\varphi\geq\tau\max(1,4\rho_s/\rho_c)$ and $\lambda\geq\max\big(\frac{\rho^2_c+\tau^2b}{\tau\rho^2_c},\varphi\delta_0\big)$; once substituted into \eqref{eq:smoothness_implies}, proves \eqref{eq:FL_Convergence}.
\end{IEEEproof}
\smallskip

Theorem \ref{thm:FL_Convergence} rigorously bounds the difference in the objective value of the optimal model $\vw^{\rm opt}$ and the one learned by \ac{cpa} over $t$ learning rounds with \lsgd; i.e., the users' batch size is set to $1$.
By taking $t$ to be asymptotically large in \eqref{eq:FL_Convergence}, we obtain the asymptotic convergence profile, indicating that \ac{cpa} with \lsgd converges at a rate of $\cO(1/t)$.
This reverse dependence on $t$ was formed due to the specific design of the step size $\eta_t$ to gradually decreases, which is also known to contribute to the convergence of \ac{fl} \cite{stich2018local, li2019convergence}, and the usage of a lattice with a gradually decaying dynamic range to fit the quantizer to the decaying magnitude of the model updates expected for converging \ac{fl}. The asymptotic rate of $\cO(1/t)$ is of the same order of convergence as \ac{fl} with neither privacy nor compression constraints \cite{stich2018local, li2019convergence}, indicating the ability of \ac{cpa} to satisfy these requirements while mitigating their harmful effects on the learning procedure. 

In the non-asymptotic regime, the integration of compression and privacy techniques does influence model convergence, as revealed in Theorem~\ref{thm:FL_Convergence} by the coefficient $\ConvCoeff$. The scalability of \ac{cpa} is reflected in the first summand of \eqref{eq:FL_convergence_b_term}, which vanishes as the number of users $K$ grows. The terms which do not vanish as $K\rightarrow \infty$, i.e., the last two summands in \eqref{eq:FL_convergence_b_term}, stem from the usage of multiple local iterations per round and from the presence of statistical heterogeneity, respectively~\cite{li2019convergence}; both are common properties of \ac{fl} that are not targeted in our design of \ac{cpa}.
 
\section{Experimental Study}\label{sec:experiments}
In this section we numerically evaluate \ac{cpa} and compare it to alternative approaches for compression and privacy in \ac{fl}. We consider the federated training of different model architectures for handwritten digit identification with the MNIST dataset as well as image classification based on CIFAR-$10$. We quantify the distortion induced by \ac{cpa}, the accuracy of the learned models, and the robustness to malicious users\footnote{The source code used in our experimental study, including all the hyper-parameters, is available online at \url{https://github.com/langnatalie/CPA}.}.

\subsection{Setup}\label{subsec:setup}
We consider \ac{fl} using \lsgd with the number of edge users varying from as small as $K=10$ to massive networks with $K=1000$, each studying different aspects in the design of \ac{cpa}. 

\subsubsection*{Baselines}
We numerically evaluate the following schemes:
\begin{description} 
    \item[{\em vanilla \ac{fl}}:] assuring neither privacy nor compression.     
    \item[{\em \ac{cpa}}:]  1-bit \ac{cpa} with a scalar quantizer; 
    \item[{\em \ac{cpa} w/o \ac{rr}}:]  1-bit \ac{cpa} with a scalar quantizer without \ac{ldp} constraints, i.e., $\varepsilon\rightarrow\infty$; 
    \item[{\em nested \ac{cpa}}:] two-stage nested \ac{cpa} (see Subsection~\ref{subsec:nested_cpa}) with $R_{\rm c}=1$, and $R_{\rm n}=3$; 
    \item[{\em Laplace}:] local updates perturbated by a Laplacian \ac{ppn}, realize the \lm \cite{dwork2016calibrating} and satisfy only privacy.
    \item[{\em \ssgd \& \ac{rr}}:] the common {\em \ssgd} \cite{bernstein2018signsgd}, which also utilizes 1-bit representations, followed by {\em \ac{rr}}, realizing a straightforward separated design satisfying \ref{itm:ldp}-\ref{itm:scalability}.  
    \item[{\em JoPEQ}:] the scheme of \cite{lang2022joint}, which transforms randomized lattice quantization distortion   into \ac{ppn},  tackling \ref{itm:ldp}-\ref{itm:universal}.  
    \item[{\em MVU}:] the scheme of \cite{chaudhuri2022privacy}, which introduces discrete-valued \ac{ldp}-preserving perturbation to the quantized representation of the model update.
\end{description}
Unless stated otherwise, all benchmarks holding $\varepsilon$-\ac{ldp} set $\varepsilon=0.5$. As for the ones 
involving compression, they utilize a mid-tread uniform scalar quantizer, i.e., $L=1$ with bit-rate $R=1$. 
Note that by Proposition~\ref{pro:kano}, the \ac{cpa} schemes  satisfy \kano with $k=2^{LR-1}$, e.g., $k=4$ for nested \ac{cpa} with $R_{\rm n}=3$.

\subsubsection*{Evaluation Metrics}
We aim to numerically validating that \ac{cpa} indeed minimizes the excess distortion compared to individual compression and privacy enhancement operating with the same \ref{itm:ldp}-\ref{itm:universal}. 
To this end, we evaluate the observed \ac{snr} [dB] of the weights obtained by \ac{cpa} compared to the desired \ac{fa}, which we compute as the estimated variance of the model weights and divide it by the estimated variance of the distortion, namely, 
\begin{equation}\label{eq:snr_definition}
    \text{SNR}\triangleq\var(\globalFA)/\var(\globalFA - \globalCPA).
\end{equation}
Then, we compute performance scoring in terms of  both validation set and test set accuracy [\%].

\subsubsection*{Architectures}
We consider the following models in training:
\begin{description}
    \item[{\em Linear}:] the model comprised of a tunable weight matrix and a bias vector of corresponding dimensions as those of the data;  
    \item[{\em \acs{mlp}}:] a \ac{mlp} with two hidden layers and intermediate ReLU activations; 
    \item[{\em \acs{cnn}$2/3$}:] a \ac{cnn} composed of two or three convolutional layers, respectively, followed by fully connected ones, with intermediate ReLU activations, max-pooling and dropout layers.    
    \item[{\em ResNet-$18$}:] the $18$ layers deep \ac{cnn} of \cite{he2016deep}. Here, we set the initial weights to be the pre-trained version of the network trained on more than a million images from the ImageNet database, having the network able to classify images into $1000$ object categories. Therefore, we extend the model to constitute another final linear layer that maps the output into the required number of labels in accordance with the used dataset (e.g, $100$ for CIFAR-$100$).    
\end{description}

\subsubsection*{Datasets}
The above architectures are trained using the following datasets:
\begin{description}
    \item[{\em MNIST}:] the dataset resembles a handwritten digit identification task, is comprised of $28 \times 28$ gray-scale images divided into $60,000$ training examples and $10,000$ test examples; where each edge user possesses $5$ uniformly drawn samples.
    \item[{\em CIFAR-$10$}:] a natural image classification dataset,  comprised of $32 \times 32$ RGB images divided into $50,000$ training examples and $10,000$ test examples, uniformly distributed among $K$ users. 
    \item[{\em CIFAR-$100$}:]  a natural image classification dataset, consisting of $60,000$ color images partitioned into $100$ classes, with each class holding $600$ images. The dataset is further divided into $50,000$ training images and $10,000$ testing image, uniformly distributed among $K$ users. 
\end{description}

\subsection{Results}\label{subsec:exp_results}
\subsubsection*{Performance}
\begin{figure}
    \centering
    \includegraphics[width=0.9\columnwidth]{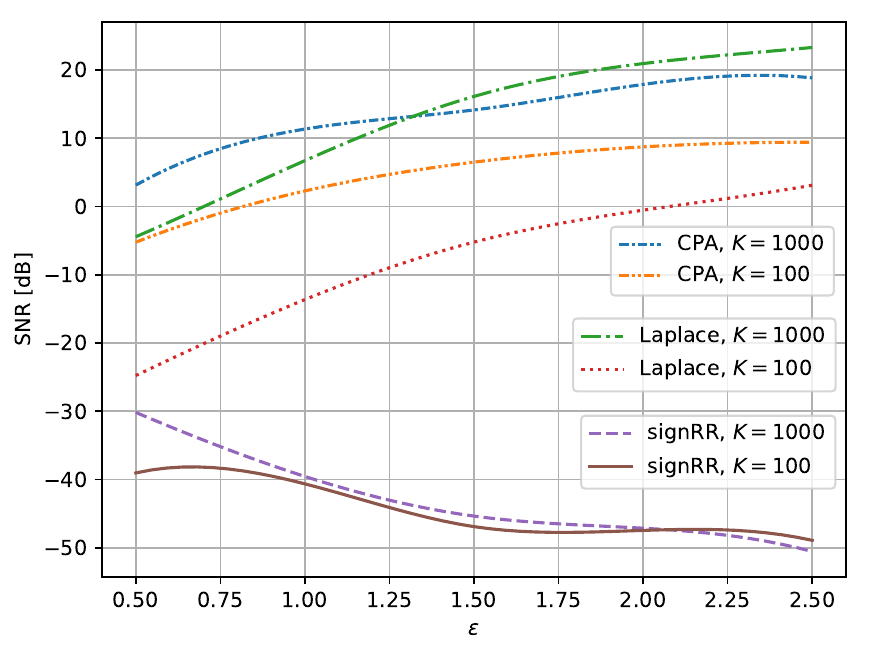}
    \caption{\ac{snr} versus $\varepsilon$ in the received models training a linear regression model using the MNIST dataset for $K\in\{100,1000\}$ edge users.}
    \label{fig:journal_SNR_1000and100_users}
\end{figure}
\begin{figure}
    \centering
    \includegraphics[width=0.9\columnwidth]{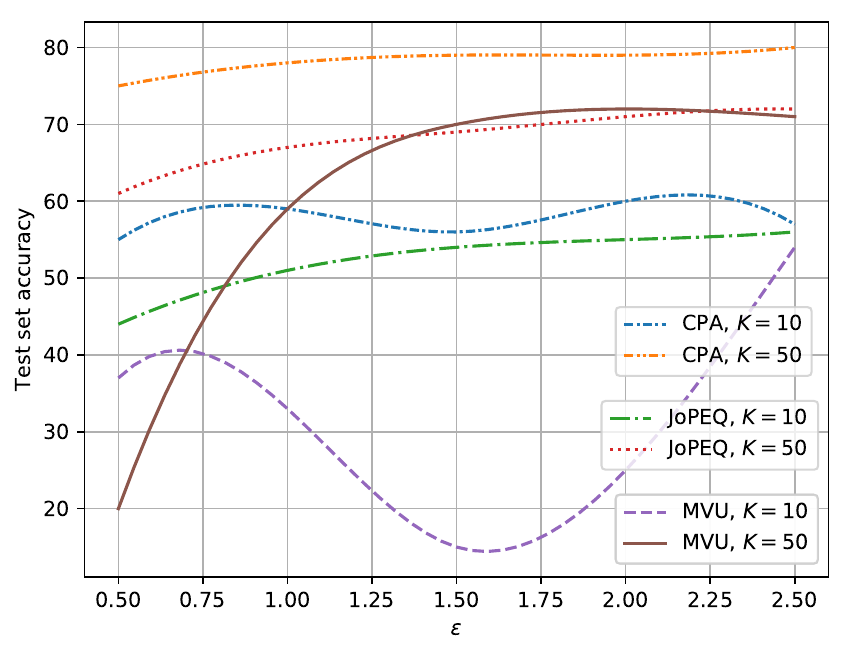}
    \caption{Test set accuracy versus $\varepsilon$ in the received models training a linear regression model using the MNIST dataset for $K\in\{10,50\}$ edge users.}
    \label{fig:eps_vs_test_acc}
\end{figure}

We begin by considering the \ac{fl} training of a linear regression model using MNIST, with $K=\{100,1000\}$ participating edge users, for different privacy budgets, i.e., $\varepsilon$ values. Accordingly, Fig.~\ref{fig:journal_SNR_1000and100_users} reports the \ac{snr} \eqref{eq:snr_definition} values as a function of the privacy budget $\epsilon$.
Evidently, \ac{cpa} attains fairly equivalent performance compared to the alternative of Laplace, which only meets
\ref{itm:ldp} and \ref{itm:universal}, while satisfying \ref{itm:ldp}-\ref{itm:scalability} altogether.
The \ac{snr} of both schemes grows with looser privacy constraints and/or more users participating; while \ssgd \& \ac{rr} demonstrates neither. This can be attributed to the coarse sign operation, whose distortion is so dominant such that it is sometimes reduced by privacy, and is barely influenced by the number of edge users taking part in the \ac{fl} training.

Following Fig.~\ref{fig:journal_SNR_1000and100_users}, we continue with measuring and monitoring the performance for variable privacy budgets not only in terms of \ac{snr} but also for test set accuracy values. For that aim, we trained a linear model on the MNIST dataset and depict in Fig.~\ref{fig:eps_vs_test_acc} the performance of the counterparts baselines: \ac{cpa}, \ac{jopeq}~\cite{lang2022joint}, and \ac{mvu}~\cite{chaudhuri2022privacy}. It in noted that both \ac{jopeq} and \ac{mvu} are not tailored for a massive number of \ac{fl} participants, and we therefore experimented $K\in\{10,50\}$. 
The expected behavior is a monotonically increasing one with incrementing $\varepsilon$, as higher budgets are attributed with lower added noise to the learning process. This is observed for all baselines for $K=50$, with lesser stability for $K=10$ (which is most notable for \ac{mvu}). For either of the values of $K$ and $\varepsilon$, \ac{cpa} performs the best, which adds to its inherent ability to benefit from operating with a massive number of users.

We proceed to numerically evaluate the effect of the parameters of \ac{cpa}, as identified in Theorems~\ref{thm:weights_distortion_bound}-\ref{thm:FL_Convergence}. For that, we inspect the  behavior of the \ac{mse} $\frac{1}{M}{\big\|\globalCPA_{t+\tau}\!-\!\globalFA_{t+\tau}\big\|}^2$ bounded in Theorem~\ref{thm:weights_distortion_bound}, in terms of both parameters dependence and relation to model accuracy; where the latter is highly correlated with the convergence bound captured in Theorem~\ref{thm:FL_Convergence}. 
For that aim, we trained a \acs{cnn}2 model on the MNIST dataset along $150$ global rounds, and evaluated the performance of both vanilla \ac{fl} and \ac{cpa}. Specifically, the latter is the $1$-bit \ac{cpa} scheme which uses a mid-tread uniform scalar quantizer, i.e., $L=1$, with bit-rate $R = 1$. 

As the relation between different bit-rates and the converged model accuracy would be monitored in the sequel in Fig.~\ref{fig:eps_R_curves}, we focus here on the impact of $K$ and $\varepsilon$.
At first, we varied the number of edge devices $K$ and fixed the privacy budget $\varepsilon=0.5$, and computed the test accuracy of the baselines compared to the \ac{mse}; as  summarized in Table~\ref{tbl:simulating_weight_distoration_bound_vs_K}.
Secondly, we repeated this procedure for fixed $K=100$ and varying $\varepsilon \ \left(p=\frac{e^\varepsilon}{1+e^\varepsilon}\right)$ in Table~\ref{tbl:simulating_weight_distoration_bound_vs_p}. 

It is revealed that the \ac{mse} indeed deceases for either of the parameters $K$ or $p$, in line with Theorem~\ref{thm:weights_distortion_bound}. Additionally, as desired, the \ac{mse} values corresponds to the model accuracy score.
In practice, this quantitatively shows that convergence in the weights guarantees convergence in model performance. It follows since we first show that despite incorporating, using the algorithm of \ac{cpa}, privacy and compression into \ac{fl}, we obtain close resemblance in weights (via the \ac{mse} metric);
and then we show that this indeed leads to similar performance in terms of the task (via the test accuracy). Thus, Tables~\ref{tbl:simulating_weight_distoration_bound_vs_K}-\ref{tbl:simulating_weight_distoration_bound_vs_p} further support the relevance of Theorems~\ref{thm:weights_distortion_bound}-\ref{thm:FL_Convergence} in the analytical analysis of \ac{cpa}.

\begin{table}[t!]
\centering
\caption{Empirical evaluation for different number of \ac{fl} participants}
\begin{tabular}{|c|c|c|c|c|c|}
\hline
& \multicolumn{5}{c|}{Number of edge participants $K$} \\
\cline{2-6}
& $50$ & $100$ &  $400$ &
$700$ & $1000$\\
\hline
FedAvg, test set acc. & 94 & 94 & 94 & 95 & 95  \\
\ac{cpa}, test set acc. & 93 & 96 & 96 & 96 & 96 \\
\acs{mse} & 0.011 & 0.007 & 0.0036 & 0.0029 & 0.0031 \\
\hline
\end{tabular}
\label{tbl:simulating_weight_distoration_bound_vs_K}
\end{table} 
\begin{table}[t!]
\centering
\caption{Empirical evaluation of for different privacy budgets}
\begin{tabular}{|c|c|c|c|c|c|}
\hline
& \multicolumn{5}{c|}{\ac{ldp} budget $\varepsilon$ $\left(p=\frac{e^\varepsilon}{1+e^\varepsilon}\right)$} \\ 
\cline{2-6}
& $0.1$ & $0.15$ &  $0.2$ &
$0.25$ & $0.3$ \\
\hline
FedAvg, test set acc. & \multicolumn{5}{c|}{94}  \\
\cline{2-6}
\ac{cpa}, test set acc. & 87 & 89 & 92 & 92 & 93 \\
\acs{mse} & 0.0068 & 0.0065 & 0.0063 & 0.0062 & 0.0062 \\
\hline
\end{tabular}
\label{tbl:simulating_weight_distoration_bound_vs_p}
\end{table} 
\begin{table}[t!]
\centering
\caption{Baselines test set accuracy results}
\begin{tabular}{|c|c|c|c|c|c|}
\hline
& \multicolumn{2}{c|}{MNIST}
& \multicolumn{2}{c|}{CIFAR-$10$}\\
\cline{2-5}
& Linear & MLP &
CNN2 & CNN3 \\
\hline
vanilla FL & 87 & 90 & 48 & 60 \\
Laplace & 86 & 88 & 45 & 60 \\
\ssgd \& \ac{rr} & 79 & 10 & 10 & 10\\
JoPEQ & 82 & 78 & 47 & 66 \\
\ac{cpa} w/o \ac{rr} & 87 & 80 & 43 & 47 \\
\ac{cpa} & 85 & 86 & 50 & 67 \\
\hline
\end{tabular}
\label{tbl:summerizing_architectures}
\end{table}
\subsubsection*{Convergence}
\begin{figure}[t!]
    \centering
    \includegraphics[width=0.9\columnwidth]{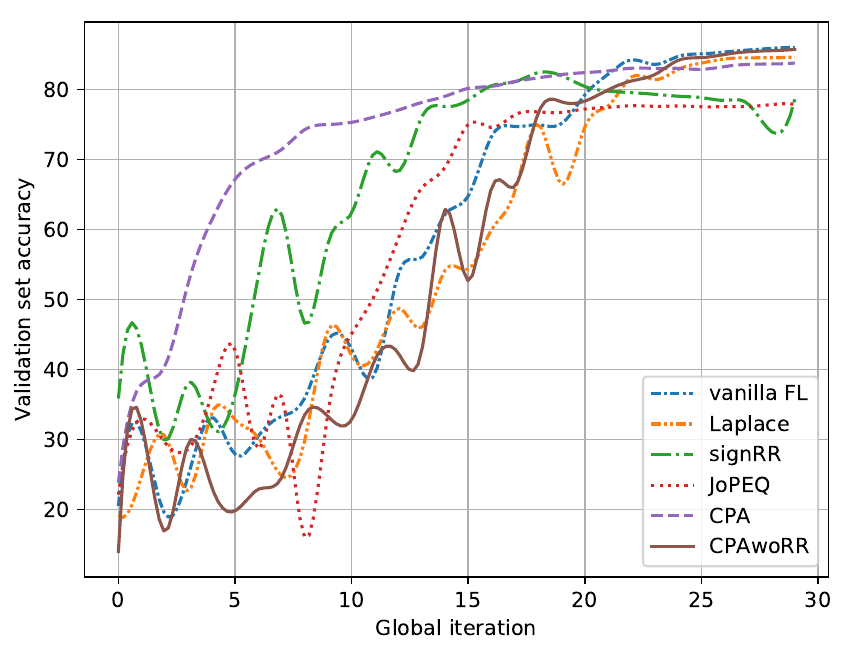}
    \caption{Convergence profile of different \ac{fl} schemes training a linear regression model using the MNIST dataset with $K=1000$ edge users.}
    \label{fig:convergence_linear_mnist}
\end{figure}

Next, we evaluate how the reduced excess distortion of \ac{cpa} is translated into an improved learning. We depict in Fig.~\ref{fig:convergence_linear_mnist} the validation set learning curves of all referenced methods. Fig.~\ref{fig:convergence_linear_mnist} indicates that \ac{cpa} performs similarly to vanilla \ac{fl} which satisfies neither privacy (\ref{itm:ldp}) nor compression (\ref{itm:rate}), while simultaneously assuring both. 
We further observe that the straightforward \ssgd \& \ac{rr} suffers from excessive distortion which deteriorates its learned model accuracy due to the usage of distinct mechanisms for quantization and privacy, as illustrated Fig.~\ref{fig:journal_SNR_1000and100_users}. A similar observation (though of a less notable gain) is noted in comparison to the joint design via \acs{jopeq} operating with the same rate of one bit per sample. 

\begin{figure}[t!]
    \centering
      \includegraphics[width=0.9\columnwidth]{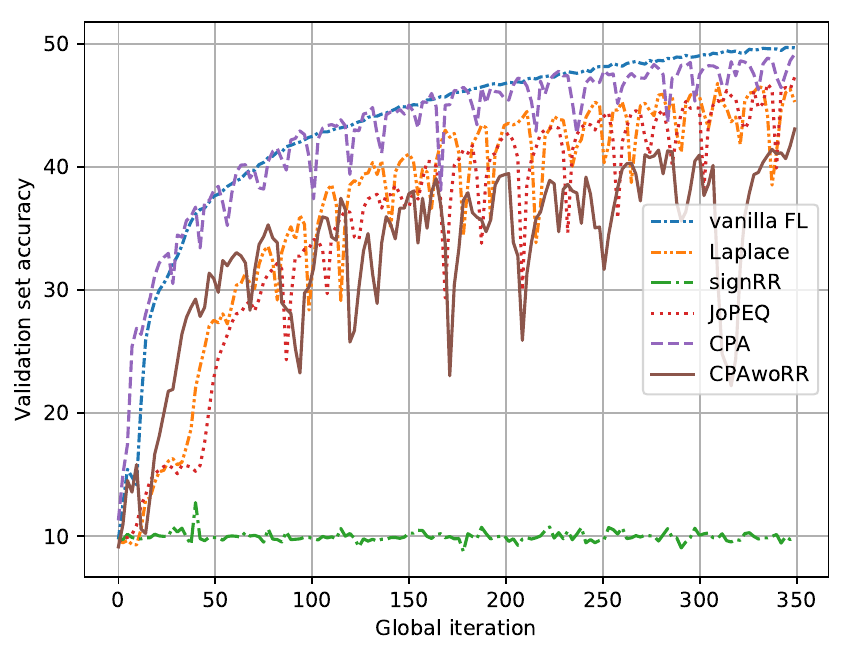}
    \caption{Convergence profile of different \ac{fl} schemes training a $2$-layered \ac{cnn} model using the CIFAR-$10$ dataset with $K=1000$ edge users.}
    \label{fig:convergence_cnn2_cifar10}
\end{figure}

We continue with showing that \ac{cpa} is beneficial regardless of the model specific design. We report in Table~\ref{tbl:summerizing_architectures} the baselines' converged models test accuracy results also for an \ac{mlp} model, 
showing in line findings with the linear model. 
Table~\ref{tbl:summerizing_architectures} also reports reports the baselines' converged models test accuracy results for two \acp{cnn} trained for the CIFAR-$10$ dataset using $K=1000$ users; while Fig.~\ref{fig:convergence_cnn2_cifar10} describes the validation set learning curves of all referenced methods.
Unlike the handwritten digit classification, in the current task it is harder for the models to converge, particularly for \ssgd \& \ac{rr}, which utterly fails to converge as revealed by Table~\ref{tbl:summerizing_architectures}. 
For the CNN3 model, we can see that the performance of both \ac{cpa} and JoPEQ is alike, yet the former holds \ref{itm:ldp}-\ref{itm:scalability} while the latter only does so for \ref{itm:ldp}-\ref{itm:universal}.

It is noted that when training deep models, adding a minor level of distortion can sometimes improve the final model performance, see, e.g., \cite{Guozhong1995NoiseBackprop,sery2021over}. Hence, \ac{cpa} without \ac{rr} does not necessarily outperform \ac{cpa} with \ac{ldp} consideration; as evidenced in Fig~\ref{fig:convergence_cnn2_cifar10} and  Table~\ref{tbl:summerizing_architectures}, having \ac{cpa} without \ac{rr} outperforms its nosier \ac{cpa} counterpart. Nevertheless, the opposite holds in Fig.~\ref{fig:convergence_linear_mnist} and Table~\ref{tbl:nested_cpa}, which consider a simpler (and shallower) linear  model. There, the perturbation induced by \ac{rr} have a consistent harmful effect on the trained model.

\begin{figure}
    \centering
    \includegraphics[width=0.9\columnwidth]{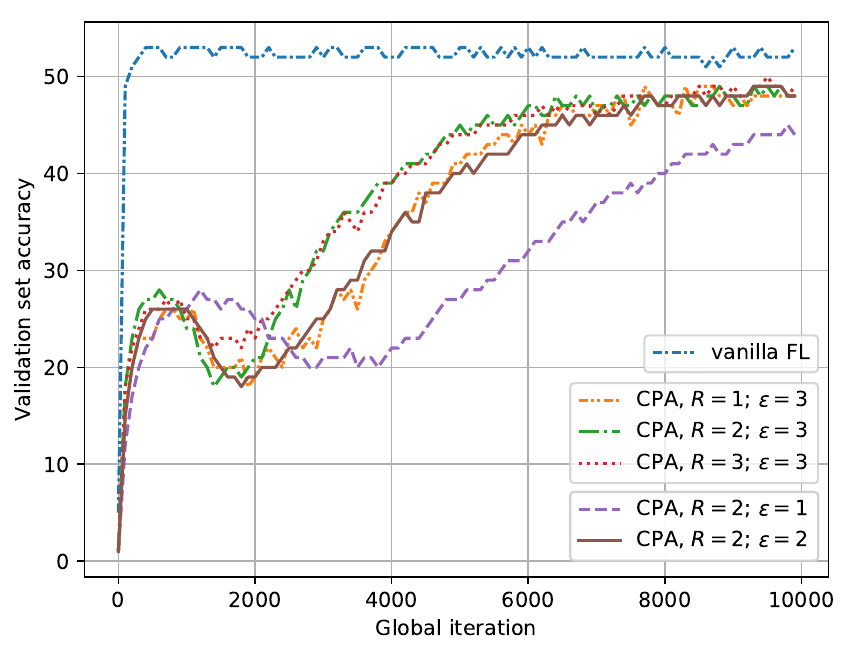}
    \caption{Convergence profile of \ac{cpa} for different $R, \ \varepsilon$ values; training a ResNet-$18$ model using the CIFAR-$100$ dataset with $K = 10$ edge users.}
    \label{fig:eps_R_curves}
\end{figure}

To further support the utility of $1$-bit \ac{cpa} regardless of the chosen dataset and/or model architecture, even for small-scale deployments; we depict in Fig.~\ref{fig:eps_R_curves} the convergence profile of \ac{cpa} training ResNet-$18$ on CIFAR-$100$ using merely $K=10$ clients. There, different combinations of the bit-rate $(R)$ and the privacy budget $(\varepsilon)$ are tested and referenced to the performance achieved with vanilla \ac{fl}, constrained with neither privacy nor compression. The training performed over $10K$ global rounds, each for $3$ local iterations, using the {\em Adam} optimizer \cite{kingma2014adam}.

It can be observed that, as expected, vanilla \ac{fl} performs best and converges fast, having no noise being added to its learning process. \ac{cpa}, on the other hand, takes longer to converge while it attains a slightly lower accuracy score, which is also attributed to the fact that only few users participate rather than hundreds and thousands of them. Furthermore,  Fig.~\ref{fig:eps_R_curves} reveals the trade-off between $R$, $\varepsilon$, and accuracy; having the privacy being the more dominant factor to deteriorate the accuracy.

\subsubsection*{Nested Operation}
We next numerically validate the gains of \ncpa. As detailed in Subsection~\ref{subsec:nested_cpa}, \ncpa alleviates the tradeoff between the number of users $K$ and the quantization rate $R$. This is because for a limited number of users, a large number of lattice points typically results in a less accurate estimation of the probability over the lattice points, which in turn translates into degraded models. 
In Table~\ref{tbl:nested_cpa} we can empirically view this behavior and the ability of the nested design to mitigate its harmful effects. 

Table~\ref{tbl:nested_cpa} summarizes the test accuracy and \ac{snr} values obtained for a linear model trained on MNIST for different number of participating clients $K$; for the baselines $1$-bit \ac{cpa} and two-stage nested \ac{cpa}, broadcasting the server $1$ and $2$ bits per sample, respectively. These are contrasted with \ac{jopeq}, which utilizes a conventional multi-bit quantizer of rate $R=2$. 
There, it is indicated that the higher the number of users is, the lessen the improvement of the nested operation over the $1$-bit scheme for both accuracy and \ac{snr} metrics. \ac{jopeq} is comparable to \ncpa in terms of bits per sample $(R)$, yet demonstrating an inferior performance, equivalent to that of single-bit \ac{cpa} for $K=10$; while the latter is far better in the large-scale scenario, what further supports its benefits in massive deployments.

\begin{table}
\centering
\caption{Test set accuracy and SNR results for 1-bit and \ncpa}
\begin{adjustbox}{width=\linewidth} 
\begin{tabular}{|c|c|c|c|c|c|c|}
\hline 
&  \multicolumn{2}{c|}{$K=10$} &
\multicolumn{2}{c|}{$K=100$} &
\multicolumn{2}{c|}{$K=1000$} \\
\cline{2-7}
& Acc. & SNR 
& Acc. & SNR 
& Acc. & SNR\\
\hline
\ac{cpa} & 49 & -17.57 & 81 & -0.07 & 85 & 0.09 \\
\ncpa & 59 & -6.16 & 83 & 0.08 & 86 & 0.17\\
\ac{jopeq} & 46 & -31.45 & 64 & -19.35 & 70 & -4.45\\
\hline
\ac{cpa} w/o \ac{rr} & 60 & -5.09 & 84 & 10.10 & 87 & 15.51 \\
\ncpa w/o \ac{rr} & 63 & 3.23 & 85 & 12.92 & 87 & 24.18 \\
\hline
\end{tabular}
\end{adjustbox}
\label{tbl:nested_cpa}
\end{table} 

\subsubsection*{Byzantine Robustness} 
We conclude by verifying \ac{cpa}'s toleration under colluding malicious participants. 
Table~\ref{tbl:malicious_users} reports the test accuracy of the converged models of the datasets and architectures considered. We simulate manipulations of a subset of the users, where it is either the scenario that a user is sending its 1-bit data constantly as $'1'$; or randomly flipping it; referenced to the result achieved with None. 
We observe that \ac{cnn}$3$ `None' does not necessarily outperform its nosier `1' or `Flip' counterparts.
This phenomenon is similar to the one evidenced in Fig.~\ref{fig:convergence_cnn2_cifar10} and Table~\ref{tbl:summerizing_architectures} for \ac{cpa} with and without \ac{rr}.
\ac{cpa}'s immunity is observed regardless of the model and/or data chosen as Table~\ref{tbl:malicious_users} indicates a degrade of single percents in accuracy under the simulated attacks. This further ensures that \ac{cpa}, in addition to being a joint compression and privacy mechanism for massive deployments, also provides robustness to Byzantine adversaries.

\begin{table}
\centering
\caption{\ac{cpa}'s test set accuracy with a subset of malicious users K=1000}
\begin{tabular}{|c|c|c|c|c|c|c|}
\hline
\multirow{3}{*}{\begin{tabular}{@{}c@{}}Malicious \\ subset\end{tabular}}
& \multicolumn{6}{c|}{MNIST} \\
\cline{2-7}
& \multicolumn{3}{c|}{Linear}
& \multicolumn{3}{c|}{MLP}\\
\cline{2-7}
& None & '1's &  Flip &
None & '1's &  Flip \\
\hline
20\% & 85 & 85 & 85 & 86 & 85 & 84 \\
30\% & 85 & 84 & 84 & 86 & 84 & 84 \\
\hline
& \multicolumn{6}{c|}{CIFAR-$10$} \\ 
\cline{2-7}
& \multicolumn{3}{c|}{CNN2}
& \multicolumn{3}{c|}{CNN3}\\
\cline{2-7}
& None & '1's &  Flip &
None & '1's &  Flip\\
\hline
20\% & 50 & 48 & 48 & 67 & 68 & 69\\
30\% & 50 & 47 & 48 & 67 & 69 & 66\\
\hline
\end{tabular}
\label{tbl:malicious_users}
\end{table}

\section{Conclusions}\label{sec:conclusions}
We proposed \ac{cpa}, which realizes quantization and privacy in scalable and robust \ac{fl}.
\ac{cpa} combines nested lattice quantization and encoding via a random codebook, with a dedicated \ac{rr} mechanism and discrete histogram aggregations to yield provable desired privacy and anonymity levels in a manner that is scalable to large networks and resilient to malicious manipulations. 
Our analysis characterizes the excess distortion induced by \ac{cpa} and its convergence, showing that 
it achieves similar asymptotic convergence profile as \ac{fl} without privacy or compression considerations.
We demonstrated that \ac{cpa} results with less distorted and more reliable models compared to alternative compression and privacy \ac{fl} methods, while approaching the performance achieved without these constraints and demolishing poisoning attacks.

\bibliographystyle{IEEEtran}
\bibliography{IEEEabrv,refs}
\vspace{-1cm}
\begin{IEEEbiography}[{\includegraphics[width=1.25in,height=1.25in,clip, angle=270,origin=c, keepaspectratio]{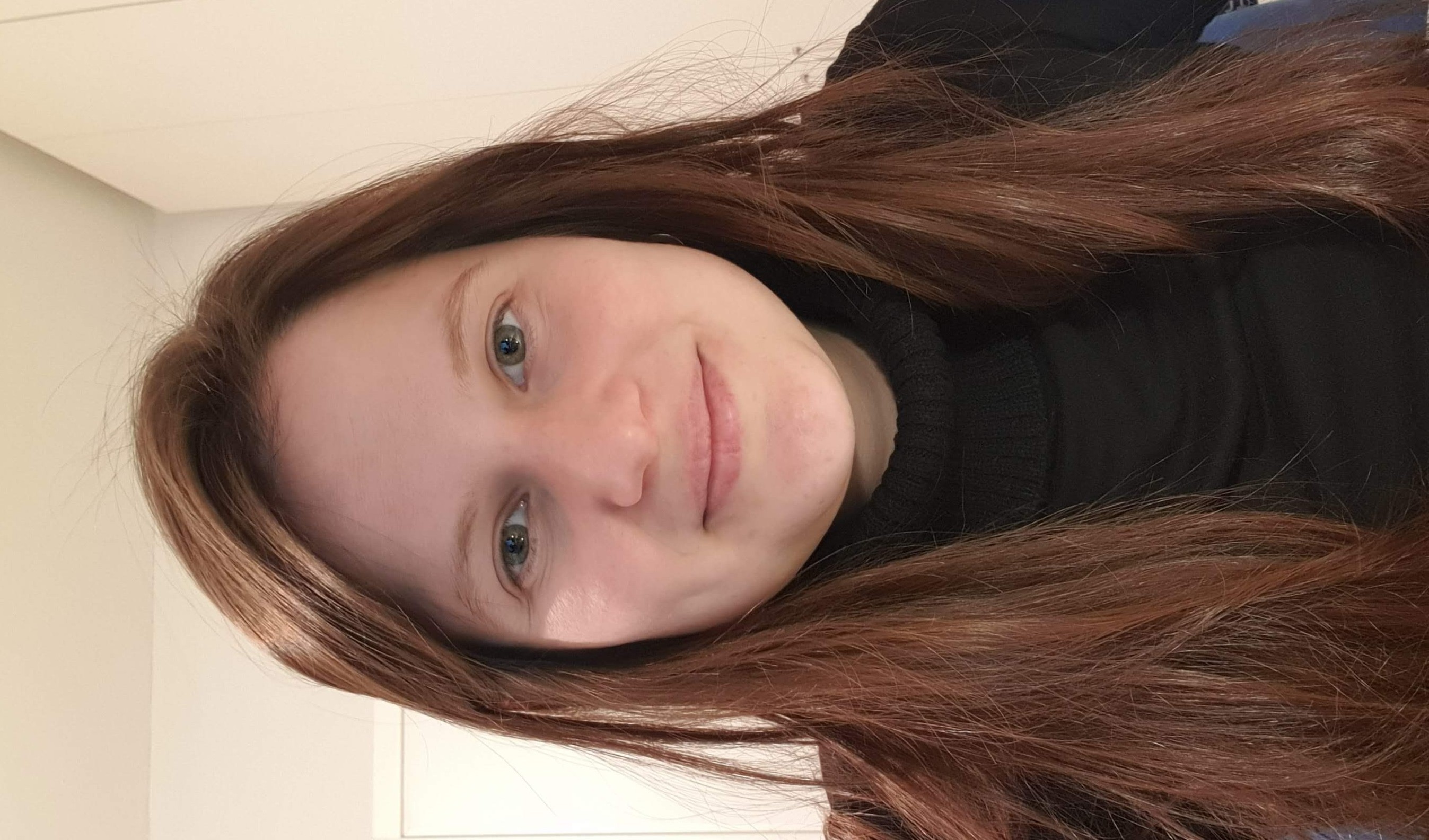}}]{Natalie Lang} received her B.Sc. and M.Sc. degrees in 2019 and 2021, respectively, from Ben-Gurion University, Israel, all in Electrical and Computer Engineering. She is currently pursuing a Ph.D. degree in Electrical and Computer Engineering at Ben-Gurion University. Her research interests include data privacy, compression, and the joint design of both for federated learning  and  signal processing. 
\end{IEEEbiography}
\vspace{-1cm}
\begin{IEEEbiography}[{\includegraphics[width=1in,height=1.25in,clip,keepaspectratio]{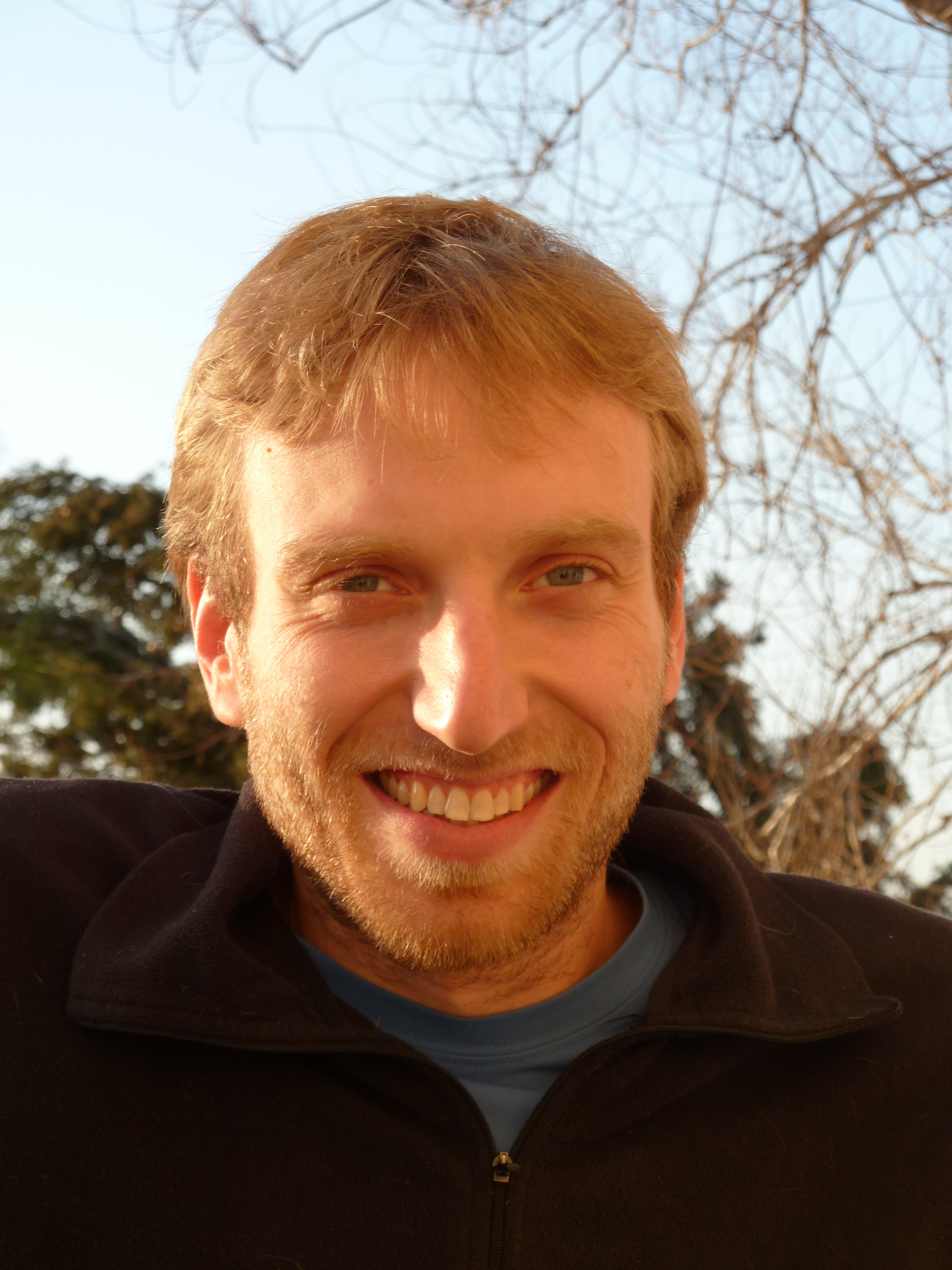}}]{Nir Shlezinger}  (M’17-SM’23) is an assistant professor in the School of Electrical and Computer Engineering at Ben-Gurion University, Israel. He received his B.Sc., M.Sc., and Ph.D. degrees in 2011, 2013, and 2017, respectively, from Ben-Gurion University, Israel, all in electrical and computer engineering. From 2017 to 2019, he was a postdoctoral researcher at the Technion, and from 2019 to 2020, he was a postdoctoral researcher at the Weizmann Institute of Science, where he was awarded the FGS Prize for his research achievements. He is the recipient of the 2024 IEEE ComSoc Fred W. Ellersick Award, and the 2024 Krill Prize for outstanding young researchers.  His research interests include communications, information theory, signal processing, and machine learning.
\end{IEEEbiography}
\vspace{-1cm}
\begin{IEEEbiography}[{\includegraphics[width=1in,height=1.25in,clip,keepaspectratio]{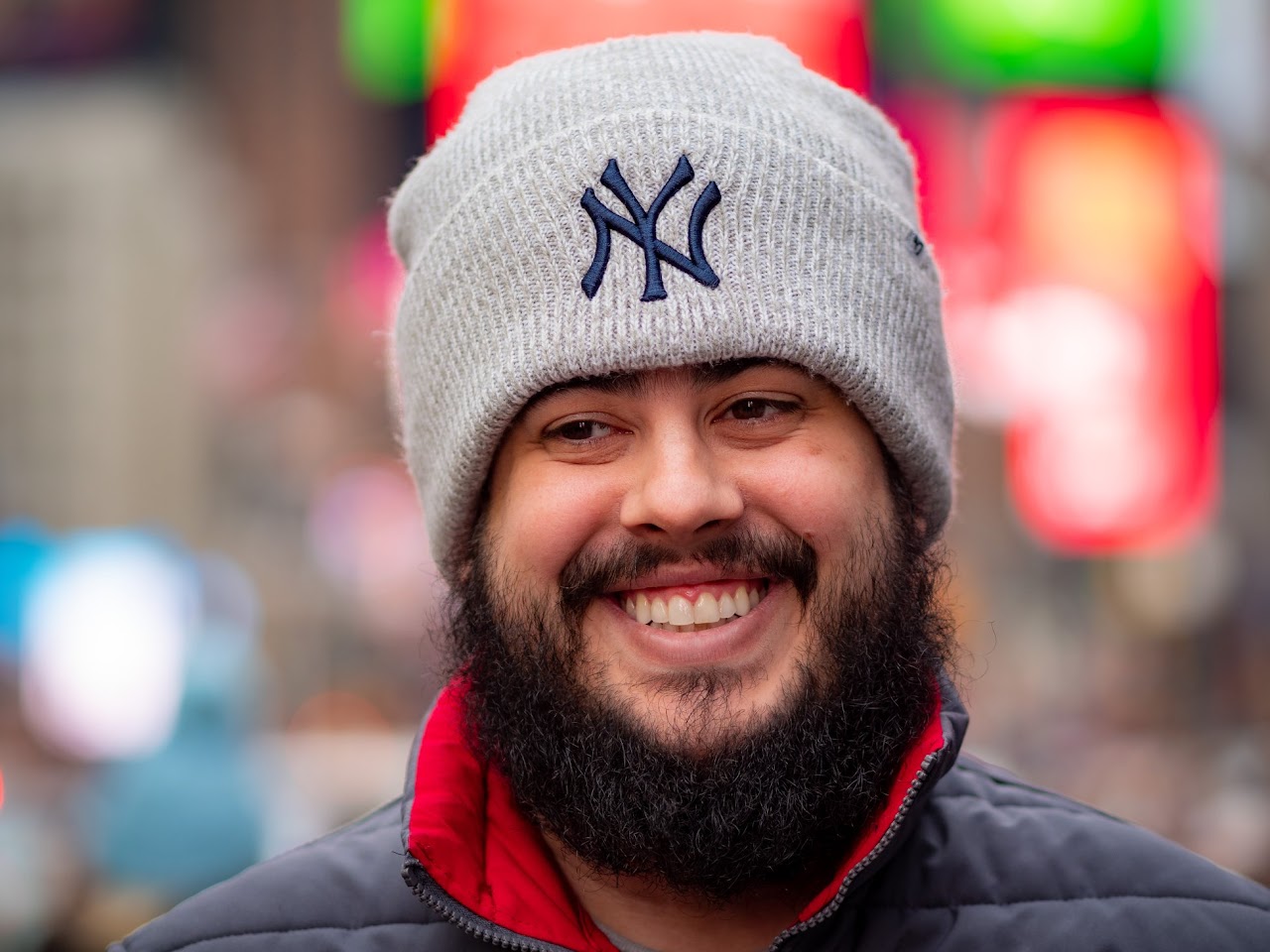}}]{Rafael G. L. D'Oliveira} is an Assistant Professor at the School of Mathematical and Statistical Sciences at Clemson University. He received a B.S. and an M.S. degree in mathematics and a Ph.D. degree in applied mathematics from the University of Campinas in Brazil in 2009, 2012, and 2017. He was a postdoctoral research associate with the Research Laboratory of Electronics at the Massachusetts Institute of Technology from 2020 to 2022, with Rutgers University from 2018 to 2019, and with the Illinois Institute of Technology in 2017. He did a research internship at Telecom Paristech from 2015 to 2016. His research interests include Privacy and Security, Distributed Computing, Coding Theory, and Information Theory.
\end{IEEEbiography}
\vspace{-1cm}
\begin{IEEEbiography}[{\includegraphics[width=1in,height=1.25in,clip,keepaspectratio]{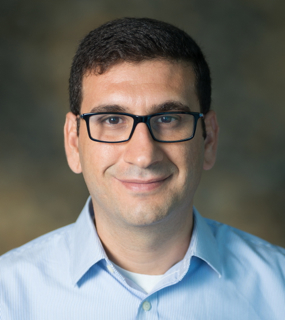}}]{Salim El Rouayheb} (Senior Member, IEEE) received the Diploma degree in electrical engineering from the Faculty of Engineering, Lebanese University, Roumieh, Lebanon, in 2002, the M.S. degree from the American University of Beirut, Lebanon, in 2004, and the Ph.D. degree in electrical engineering from Texas A$\&$M University, College Station, in 2009. He was a Post-Doctoral Research Fellow with UC Berkeley from 2010 to 2011 and a Research Scholar with Princeton University from 2012 to 2013. He was an Assistant Professor with the ECE Department, Illinois Institute of Technology, from 2013 to 2017. In 2019, he held the Walter Tyson Junior Faculty Chair at Rutgers University, where he is now an Associate Professor in the ECE Department. He received the Google Faculty Award in 2018 and the NSF CAREER Award in 2016. His research focuses on information-theoretic security and data privacy, with an emphasis on distributed learning.
\end{IEEEbiography}

\end{document}

%% file: Preamble.tex
\newif\ifproofs
\proofstrue 

\newif\ifCLASSOPTIONcompsoc
\CLASSOPTIONcompsoctrue 
\newif\ifextended

\usepackage{mathtools, amsthm, amsmath,amssymb,amsfonts, dsfont}
\usepackage{textcomp, acronym, url, comment, footnote}
\usepackage{graphicx, xcolor, enumitem, xspace, diagbox, multicol, multirow, adjustbox, fancyhdr} 
\usepackage{dirtytalk, pifont}
\usepackage[linesnumbered,ruled,vlined]{algorithm2e}
\usepackage[usestackEOL]{stackengine}

\let\oldnl\nl
\newcommand{\nonl}{\renewcommand{\nl}{\let\nl\oldnl}} 
\setlist[description]{font=\normalfont}

\ifCLASSOPTIONcompsoc
  \usepackage[nocompress]{cite}
  \usepackage[hidelinks]{hyperref}
\else
  \usepackage{cite}
  \usepackage[bookmarks, colorlinks]{hyperref}  
\fi

\newcommand{\cmark}{\ding{51}}%
\newcommand{\xmark}{\ding{55}}%

\newcommand{\user}{r} 
\newcommand{\subvec}{i} 
\newcommand{\dimension}{d} 
\newcommand{\sample}{j^\user_t} 
\newcommand{\sampleTag}{j^{\user'}_t} 

\newcommand{\virtual}{\vz} 
\newcommand{\grad}{\vg_t} 
\newcommand{\Noisygrad}{\tilde\vg_t} 

\newcommand{\weights}{\vw^\user} 
\newcommand{\weightsTag}{\vw^{\user'}} 
\newcommand{\updates}{\vh^\user_{t,\subvec}} 
\newcommand{\updatesTau}{\vh^\user_{t+\tau,\subvec}} 

\newcommand{\HerrorTau}{\eta_{t+\tau,\subvec}^l} 
\newcommand{\Herror}{\eta_{t,\subvec}^l} 
\newcommand{\QerrorTau}{\ve^\user_{t+\tau,\subvec}} 
\newcommand{\CPAerror}{\vd_{t+1}} 

\newcommand{\codeword}{\vv^\user_{t,\subvec}} 
\newcommand{\RRcodeword}{\tilde\vv^\user_{t,\subvec}} 
\newcommand{\RRhistogram}{\tilde\vv_{t,i}} 
\newcommand{\RRhistogramTau}{\tilde\vv_{t+\tau,i}} 
\newcommand{\LDPbit}{b^\user_{t,\subvec}} 
\newcommand{\bit}{\bar{b}^\user_{t,\subvec}} 

\newcommand{\Qword}{\vq^l} 
\newcommand{\nested}{\ensuremath{\cL^{\rm n}}} 
\newcommand{\coarse}{\ensuremath{\cL^{\rm c}}} 
\newcommand{\coarseP}{\ensuremath{\cP^{\rm c}_0}} 
\newcommand{\fine}{\ensuremath{\cL^{\rm f}}} 
\newcommand{\globalCPA}{\vw^{\rm CPA}} 
\newcommand{\globalFA}{\vw^{\rm FA}} 
\newcommand{\latticeM}{\bar\sigma^2_\cL} 
\newcommand{\ConvCoeff}{b}

\newcommand{\kano}{$k$-anonymity\xspace}
\newcommand{\lm}{Laplace mechanism\xspace}

\newcommand{\ssgd}{signSGD\xspace}
\newcommand{\lsgd}{local-\ac{sgd}\xspace}
\newcommand{\nq}{nested quantization\xspace}
\newcommand{\vcpa}{$1$-bit CPA\xspace} 
\newcommand{\ncpa}{nested CPA\xspace} 

\newcommand{\cA}{\mathcal{A}}
\newcommand{\cD}{\mathcal{D}}
\newcommand{\cL}{\mathcal{L}}
\newcommand{\cM}{\mathcal{M}}
\newcommand{\cP}{\mathcal{P}}
\newcommand{\cO}{\mathcal{O}}
\newcommand{\cI}{\mathcal{I}}
\newcommand{\cX}{\mathcal{X}}

\newcommand{\R}{\mathbb{R}}

\newcommand{\N}{\mathbb{N}}
\newcommand{\Z}{\mathbb{Z}}

\newcommand{\E}{\mathds{E}}
\newcommand{\proba}{\mathds{P}}

\newcommand{\vw}{\boldsymbol{w}}
\newcommand{\vG}{\boldsymbol{G}}
\newcommand{\vx}{\boldsymbol{x}}
\newcommand{\vX}{\boldsymbol{X}}
\newcommand{\vz}{\boldsymbol{z}}
\newcommand{\vl}{\boldsymbol{l}}
\newcommand{\vh}{\boldsymbol{h}}
\newcommand{\vv}{\boldsymbol{v}}
\newcommand{\vq}{\boldsymbol{q}}
\newcommand{\ve}{\boldsymbol{e}}
\newcommand{\vd}{\boldsymbol{d}}
\newcommand{\vg}{\boldsymbol{g}}
\newcommand{\vy}{\boldsymbol{y}}

\DeclareMathOperator*{\sign}{{\rm sign}}
\DeclarePairedDelimiter\abs{\lvert}{\rvert}
\DeclarePairedDelimiter\ceil{\lceil}{\rceil}
\DeclareMathOperator*{\var}{\text{Var}}
\DeclareMathOperator*{\argmin}{arg\,min}

\acrodef{ml}[ML]{machine learning}
\acrodef{fl}[FL]{federated learning}
\acrodef{dq}[DQ]{dithered quantization}
\acrodef{sdq}[SDQ]{subtractive dithered quantization}
\acrodef{ldp}[LDP]{local differential privacy}
\acrodef{dp}[DP]{differential privacy}
\acrodef{sgd}[SGD]{stochastic gradient descent}
\acrodef{fa}[FedAvg]{federated averaging} 
\acrodef{ppn}[PPN]{privacy preserving noise} 
\acrodef{dlg}[DLG]{deep leakage from gradients}
\acrodef{idlg}[iDLG]{improved deep leakage from gradients}
\acrodef{cpa}[CPA]{compressed private aggregation}
\acrodef{rr}[RR]{randomized response}
\acrodef{pmga}[PMGA]{private multi-group aggregation}
\acrodef{ai}[AI]{artificial intelligence}
\acrodef{snr}[SNR]{signal-to-noise ratio}
\acrodef{mse}[MSE]{mean squared error}
\acrodef{mlp}[MLP]{multi-layer perceptron}
\acrodef{cnn}[CNN]{convolutional neural network}
\acrodef{jopeq}[JoPEQ]{joint privacy enhancement and quantization} 
\acrodef{rhs}[RHS]{right hand side} 
\acrodef{mvu}[MVU]{minimum variance unbiased}

\newtheorem{theorem}{Theorem}[section]
\newtheorem{definition}[theorem]{Definition}
\newtheorem{corollary}{Corollary}
\newtheorem{proposition}{Proposition}

\newtheorem{claim}{Claim}
\newtheorem{lemma}{Lemma}